\documentclass[10pt]{iopart}

\usepackage{iopams}  
\usepackage{amstext}

\eqnobysec 

\usepackage{graphicx}

\usepackage{hyperref}

\begin{document}

\title[Sustainability, collapse and oscillations in a simple World-Earth model]{Sustainability, collapse and oscillations of global climate, population and economy in a simple World-Earth model}
\author{Jan Nitzbon$^{1,2,3}$, Jobst Heitzig$^{2}$, Ulrich Parlitz$^{1,4}$}

\address{$^{1}$Institute for Nonlinear Dynamics, Faculty of Physics, University of G\"ottingen, Friedrich-Hund-Platz 1, 37077 G\"ottingen, Germany}
\address{$^{2}$Potsdam Institute for Climate Impact Research, P.O. Box 601203, 14412 Potsdam, Germany}
\address{$^{3}$Alfred Wegener Institute for Polar and Marine Research, P.O. Box 600149, 14401 Potsdam, Germany}
\address{$^{4}$Max Planck Institute for Dynamics and Self-Organization, P.O. Box 2853, 37018 G\"ottingen, Germany}

\ead{\mailto{jan.nitzbon@awi.de}, \mailto{heitzig@pik-potsdam.de}}


\begin{abstract}

The Anthropocene is characterized by close interdependencies between the natural Earth system and the global human society, posing novel challenges to model development.
Here we present a conceptual model describing the long-term co-evolution of natural and socio-economic subsystems of Earth.
While the climate is represented via a global carbon cycle, we use economic concepts to model socio-metabolic flows of biomass and fossil fuels between nature and society. 
A well-being-dependent parametrization of fertility and mortality governs human population dynamics.

Our analysis focuses on assessing possible asymptotic states of the Earth system for a qualitative understanding of its complex dynamics rather than quantitative predictions.
Low dimension and simple equations enable a parameter-space analysis allowing us to identify preconditions of several asymptotic states and hence fates of humanity and planet. 
These include a sustainable co-evolution of nature and society, a global collapse and everlasting oscillations.

We consider different scenarios corresponding to different socio-cultural stages of human history.
The necessity of accounting for the ``human factor'' in Earth system models is highlighted by the finding that carbon stocks during the past centuries evolved opposing to what would ``naturally'' be expected on a planet without humans.
The intensity of biomass use and the contribution of ecosystem services to human well-being are found to be crucial determinants of the asymptotic state in a (pre-industrial) biomass-only scenario without capital accumulation. 
The capitalistic, fossil-based scenario reveals that trajectories with fundamentally different asymptotic states might still be almost indistinguishable during even a centuries-long transient phase.
Given current human population levels, our study also supports the claim that besides reducing the global demand for energy, 
only the extensive use of renewable energies may pave the way into a sustainable future.

\end{abstract}

%

\submitto{\ERL}

\maketitle
%
\ioptwocol

\clearpage
\section{Introduction}

The 
impacts humankind exerts on 
nature
on a planetary scale have become so grave that an entirely new geological epoch -- the Anthropocene -- has been proclaimed \cite{Crutzen2002},
characterized by strong 
nature-society interrelations.
Independent of whether the Anthropocene indeed depicts a novel geological epoch or not \cite{Zalasiewicz2011, Malm2014, Lewis2015, Waters2016}, 
predicting
Earth's future 
with models 
necessitates recognizing the influences humans exert on it and vice versa.
This qualitatively new relation between humans and nature poses a huge challenge for the development of suitable models, 
demanding 
a balanced representation of both the \textit{natural sphere} (ecosphere, ``Earth'') and the \textit{human sphere} (anthroposphere, ``World'') 
and 
a holistic system's perspective \cite{Schellnhuber1998, Schellnhuber1999, Verburg2016, vanVuuren2016}.
Many models of the natural Earth system (e.g., general circulation models (GCMs) or Earth system models of intermediate complexity (EMICs)) include human impacts only as an exogenous driver,
e.g., in the form of emission scenarios \cite{Moss2008}.
Integrated assessment models (IAMs) on the other hand try to simulate and/or optimize the future economic evolution 
under changing environmental conditions on 
multiple decades \cite{Weyant1996}.
However, 
only few modelling attempts 
aim
at a balanced representation of natural and socio-economic dynamics on centennial to millennial time scales \cite{Meadows1972,Meadows2004,Boumans2002,Motesharrei2014,Kittel2017}.
Conceptual World-Earth models
like the one presented 
here
try to fill 
this gap 
in the model landscape 
and thereby contribute to 
modelling the Anthrophocene.

Complementary to the development of useful models of World-Earth {\em dynamics} 
stands the challenge to 
identify
a {\em desirable condition} of the World-Earth system. 
The concept of Planetary Boundaries is a major advance in this direction regarding the natural dimension 
\cite{Rockstrom2009, Rockstrom2009a, Steffen2015}. It states that during the holocene several aggregate indicators of the Earth's state stayed within certain limits which 
define a kind of ``safe operating space'' to which humanity is adapted and which should 
not be transgressed.
Within the framework of the ``Oxfam doughnut'' these 
bounds
are supplemented by quantitative indicators of socio-economic aspects of the world, called ``social foundations'', which together are thus interpreted to define a ``safe and just operating space'' \cite{Raworth2012},
see also the
Sustainable Development Goals \cite{UnitedNations2015,Folke2016}.
The state space topology and dilemmas resulting from such boundaries can be analysed if the models are not too complex \cite{Heitzig2016}.  

A deeper qualitative understanding of the World-Earth system is facilitated by the use of rather simple, conceptual models which allow also for analytical analyses.
Examples for such attempts comprise the studies of rather local models of
natural resources co-evolving with social or population dynamics \cite{Brander1998, Brandt2013, Wiedermann2015, Barfuss2016}, but also models which address social stratification \cite{Motesharrei2014} and conceptual models on a global scale \cite{Kellie-Smith2011, Milik1996}.

Our goal here is to contribute to the latter strand of literature a simple conceptual model focussing on a few globally aggregated quantities of the natural and socio-economic subsystems 
that appear most essential to assess the desirability of the system state in terms of population, well-being, and biosphere integrity. 
As well-being and biosphere integrity depend crucially on climate and natural resource use,
our World-Earth model 
describes the temporal evolution  
of the global carbon cycle, human population, and the competition between the major energy sources, biomass and fossil fuels,
on centennial to millennial time-scales.
A particular objective of this study is to characterize the possible asymptotic paths the world could have taken, 
and to identify model parameters crucial for switching between these qualitatively different dynamic regimes.
To be able to apply the necessary techniques from dynamical systems theory, e.g., bifurcation analysis, we keep the dimension low, using only five dynamic variables, and the equations simple.

Despite this simplicity, the model is capable of qualitatively reflecting the actual dynamics seen during different stages in human history, in particular the Holocene and the Anthropocene.
We achieve this by a novel combination of a carbon cycle with well-being-driven population dynamics and economic production based on 
energy and accumulated capital.
In our model the 
global carbon cycle 
is modelled 
similar to 
\cite{Anderies2013,Heck2016}
We combine this with established concepts from economics and 
population dynamics 
and show that without such an anthroposphere component the model behaviour would deviate drastically from what is observed.

The 
paper is structured as follows:
After introducing the full model in section\,\ref{sec:model}, 
we analyse 
special cases of growing complexity that roughly relate to different eras in 
human
history in section\,\ref{sec:results}
before concluding 
in section\,\ref{sec:conclusions}. 
The appendix contains details regarding the derivation of the model, the estimation of its parameters, its bifurcation analysis, and conditions for phases of superexponential growth.

\section{Model}
\label{sec:model}

\begin{figure*}
	\centering
    \includegraphics[width=0.8\textwidth]{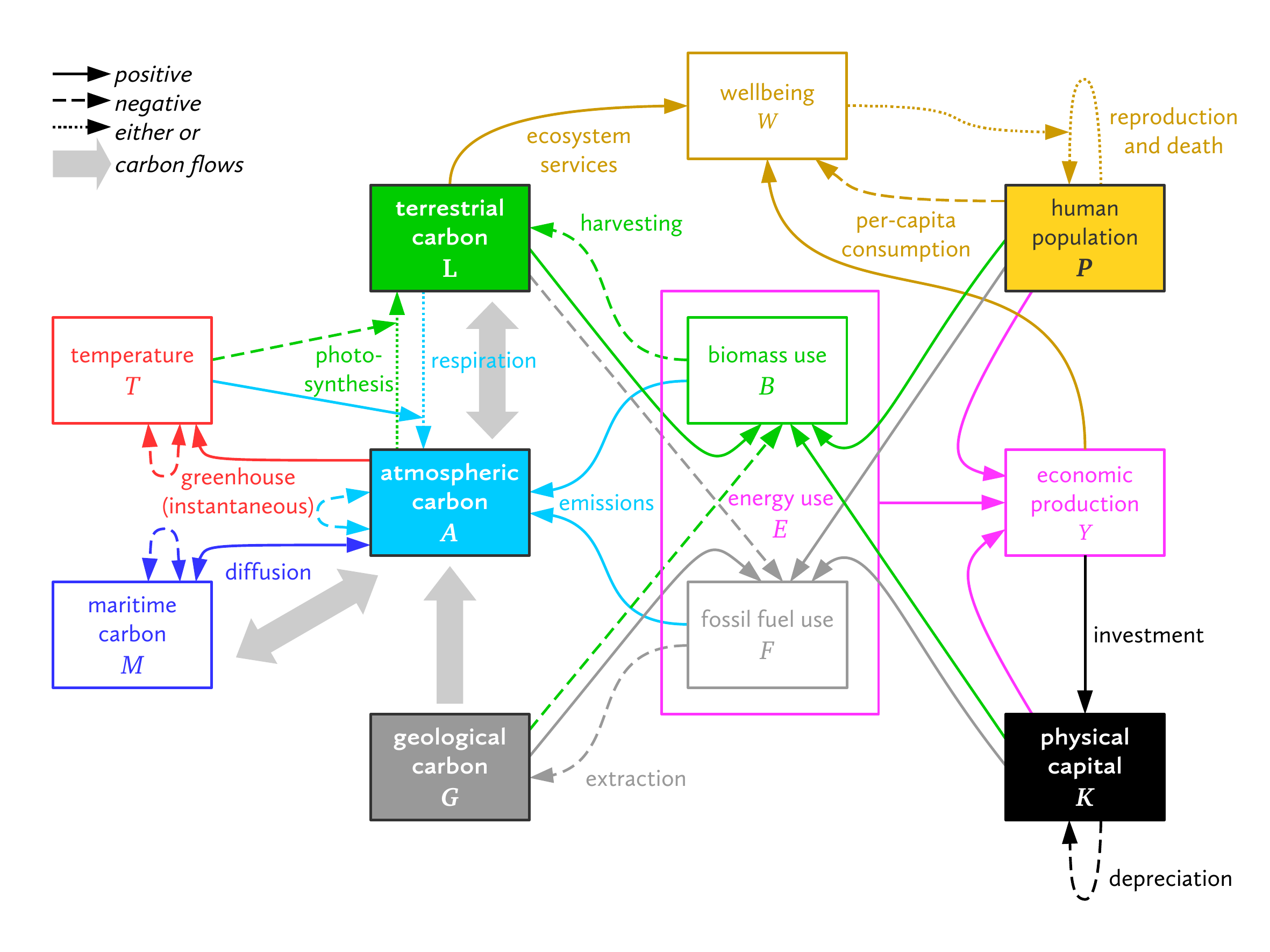}
    \caption{Overview of the model structure with five state variables (colored boxes) and several derived variables (white boxes). Arrows represent coupling processes between the variables. The left part represents the natural subsystem of the Earth (Ecosphere) via the global carbon cycle, while the right part represents socio-economic entities related to human activities in the World (Anthroposphere).}
    \label{fig:model}
\end{figure*}

\begin{table*}[!ht]
	\caption{Overview of the model parameters, their physical dimensions and the best estimate based on real-world data.}
	\label{tbl:parms}
	\begin{indented}
	\centering
	\item[]
	\begin{tabular}{lllr}
		\br
		Symbol				& Description								& Unit (H = humans)					& Estimate		\\
		\br
		$\Sigma$			& available Earth surface area				& km$^2$							& $1.5\cdot10^8$	\\
		$C^\ast$			& total available carbon stock				& GtC								& $5500$			\\
		$a_0$				& respiration baseline coefficient			& a$^{-1}$							& $0.0298$ 			\\
		$a_T$				& respiration sensitivity to temperature	& km$^2\,$a$^{-1}\,$GtC$^{-1}$ 		& $3200$ 			\\
		$l_0$				& photosynthesis baseline coefficient		& km$\,$a$^{-1}\,$GtC$^{-1/2}$		& $26.4$ 			\\
		$l_T$				& photosynthesis sensitivity to temperature	& km$^3\,$a$^{-1}\,$GtC$^{-3/2}$	& $1.1\cdot10^6$ 	\\
		$d$			& diffusion rate							& a$^{-1}$							& $0.01$ 			\\
		$m$					& solubility coefficient					& 1									& $1.5$ 			\\
		\mr
		$p$					& fertility maximum							& a$^{-1}$							& $0.04$			\\
		$W_P$				& fertility saturation well-being			& \$\,a$^{-1}\,$H$^{-1}$				& $2000$			\\
		$q$					& mortality baseline coefficient			& \$\,a$^{-2}\,$H$^{-1}$				& $20$				\\
		\mr
		$i$					& investment ratio							& 1									& $0.25$ 			 \\
		$k$					& capital depreciation rate					& a$^{-1}$							& $0.1$				 \\
		\mr	
		$a_B$				& biomass sector productivity				& GJ$^5\,$a$^{-5}\,$GtC$^{-2}\,$\$$^{-2}\,$H$^{-2}$ & varied	 \\
		$a_F$				& fossil fuel sector productivity			& GJ$^5\,$a$^{-5}\,$GtC$^{-2}\,$\$$^{-2}\,$H$^{-2}$ & varied	 \\
		$e_B$				& biomass energy density					& GJ$\,$GtC$^{-1}$					& $4\cdot10^{10}$		 \\
		$e_F$				& fossil fuel energy density 				& GJ$\,$GtC$^{-1}$					& $4\cdot10^{10}$		\\
		$y_E$				& economic output per energy input			& \$ GJ$^{-1}$						& $147$				\\
		$w_L$				& well-being sensitivity to land carbon		& \$$\,$km$^2\,$GtC$^{-1}\,$a$^{-1}\,$H$^{-1}$ 	& varied\\	
		\mr
		$C^\ast_\text{PI}$	& total pre-industrial carbon stock			& GtC								& $4000$ 			\\
		$b$					& biomass harvesting rate					& GtC$^{3/5}\,$a$^{-1}\,$H$^{-3/5}$	& $5.4\cdot10^{-7}$	\\
		$y_B$				& economic output per biomass input			& \$$\,$GtC$^{-1}$					& $2.47\cdot10^{11}$ (varied)\\
		\br
	\end{tabular}
	\end{indented}
\end{table*}

Similar to \cite{Anderies2013}, 
our conceptual model describes the 
global carbon cycle via three carbon reservoirs ---
the terrestrial ($L$, plants and soils), 
atmospheric ($A$), 
and geological ($G$) carbon stocks,
and describes the global population and economy via just two additional stocks,
human population $P$ 
and physical capital $K$ (see figure \ref{fig:model}). 
Their dynamics is governed by 
five ordinary differential equations 
\begin{eqnarray}
	\dot L &= (l_0 - l_T T) \sqrt{A/\Sigma}\, L - (a_0 + a_T T) L - B,				\label{eqn:Ldot}\\
	\dot A &= - \dot L + d (M - m A),											\label{eqn:Adot}\\
	\dot G &= - F,																	\label{eqn:Gdot}\\
	\dot P &= P \left( \frac{2 W W_P}{W^2 + W_P^2}\,p - \frac{q}{W} \right),		\label{eqn:Pdot}\\
	\dot K &= i Y - k K.															\label{eqn:Kdot}
\end{eqnarray}
The derived quantities of 
maritime carbon stock $M$,
global mean temperature 
$T$,
biomass use $B$,
fossil fuel use $F$,
economic production $Y$,
and well-being $W$
are governed by the 
algebraic equations
\begin{eqnarray}
	M &= C^\ast - L - A - G,													\label{eqn:M}\\
    T &= A/\Sigma,																\label{eqn:T}\\
    B &= \frac{a_B}{e_B}\frac{L^2 (P K)^{2/5}}{(a_B L^2 + a_F G^2)^{4/5}},		\label{eqn:B}\\
    F &= \frac{a_F}{e_F}\frac{G^2 (P K)^{2/5}}{(a_B L^2 + a_F G^2)^{4/5}},		\label{eqn:F}\\
    Y &= y_E (e_B B + e_F F),													\label{eqn:Y}\\
    W &= \frac{(1 - i) Y}{P} + w_L\frac{L}{\Sigma}.								\label{eqn:W}
\end{eqnarray}
See table~\ref{tbl:parms} and \ref{sec:par-est} for parameter meanings and estimates on the basis of available real-world data.
The three terms in $\dot L$ represent temperature-dependent photosynthesis (with atmospheric carbon fertilization) and respiration, and biomass extraction.
The second term in $\dot A$ is diffusion at the oceans' surface.
The 
terms in $\dot P$ represent well-being-dependent fertility and mortality, where fertility reaches a maximum of $p$ at $W = W_P$ and then declines again.
Finally, the 
terms in $\dot K$ are investment at a fixed savings rate and capital depreciation.
Temperature $T$ is assumed to relax instantaneously to its equilibrium value depending on $A$, 
using
a nonlinear temperature scale so it is simply proportional to $A$.
The denominator in $B$ and $F$ represents substitution effects 
in the energy sector.
Economic production $Y$ in the remaining 
sectors is 
proportional to energy input.
Well-being $W$ derives from per-capita consumption and ecosystem services (health-improving, recreational, psychological, etc.) assumed proportional to $L$.
\ref{sec:model-derivation} contains a detailed motivation and derivation of the model from physical and economic principles.

\section{Results}
\label{sec:results}

\subsection{How recent centuries' carbon cycle trends oppose purely natural dynamics}
\label{sec:natural}

\begin{figure}
	\centering
    \includegraphics[width=0.48\textwidth]{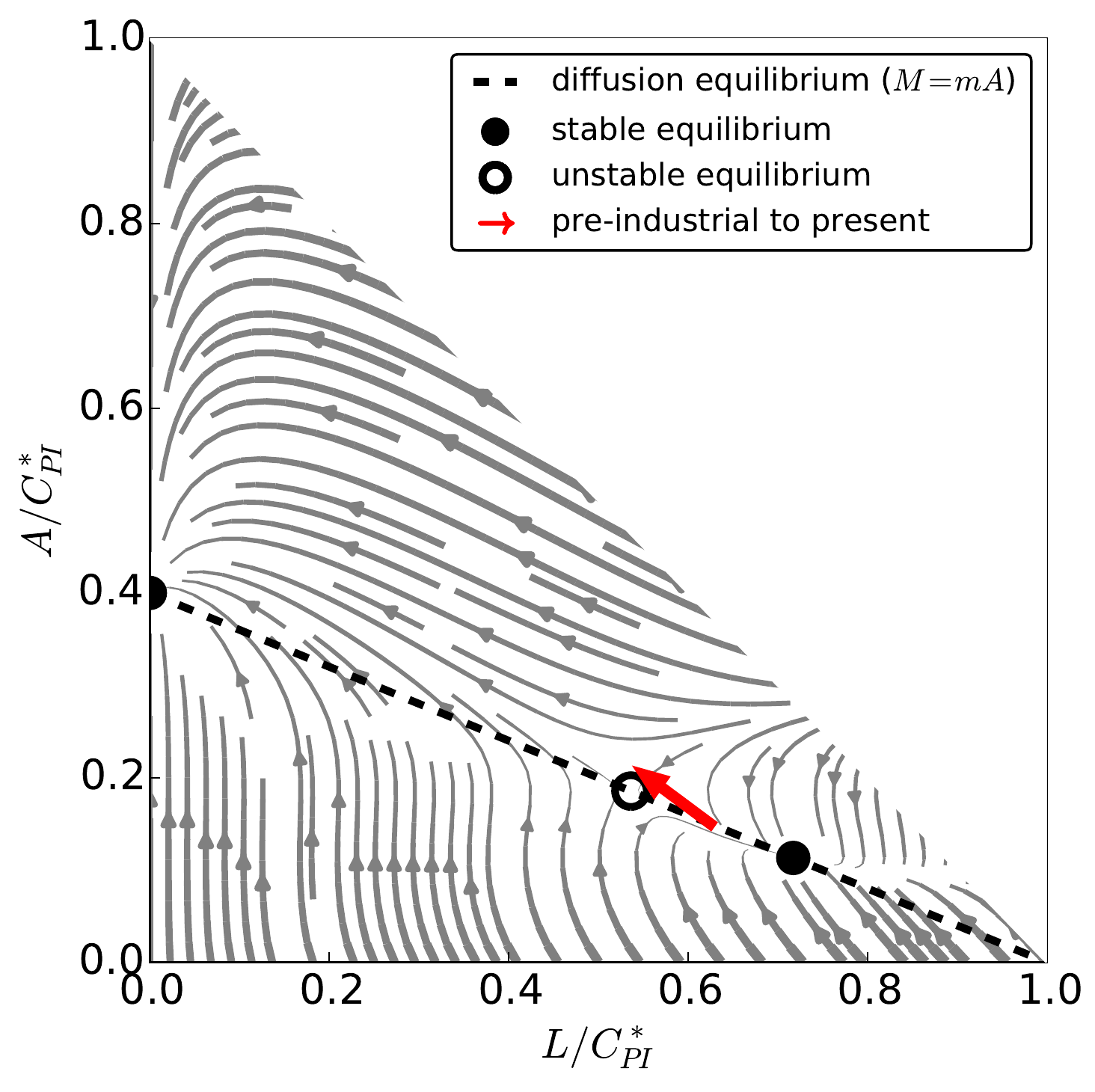}
    \caption{State space representation of the purely natural carbon cycle dynamics given by equations (\ref{eqn:Ldot}) and (\ref{eqn:Adot}) and setting $P=K=0$.
    Grey arrows show the direction of the system's evolution, thicker lines correspond to faster flow.
    On the black dashed line diffusion is in equilibrium. 
    There are three equilibria of which the ``desert'' state at $L_\text{D}=0$ and the ``forest'' state at $L_\text{F}\approx0.72 C^\ast_{PI}$ are stable.
    The red arrow reflects the actual evolution of the carbon pools from pre-industrial times until today. It opposes the natural direction of the flow, indicating the necessity of incorporating human activities into Earth system models.
    The upper right corner is not part of the state space due to the mass constraint $L+A\leq C^\ast_\text{PI}$.
    Parameters are set to the default values given in table \ref{tbl:parms}.}
    \label{fig:cGLA-phasespace}
\end{figure}

We first consider the 
natural 
carbon cycle without human interference by setting $P=K=0$.
Figure \ref{fig:cGLA-phasespace} shows the state space of the remaining two-dimensional system given by 
terrestrial ($L$) and atmospheric ($A$) carbon stocks. 
As $\dot{G}=0$, the geological carbon stock $G$ is ignored and $L$ and $A$ are normalized by the pre-industrial carbon amount of the (short-term) carbon cycle, $C^\ast_\text{PI}$.

Equilibrium states of the system require $\dot{L}=\dot{A}=0$ so that, according to equation (\ref{eqn:Adot}), 
net diffusion between the atmosphere and the upper ocean vanishes ($M=mA$). Solving (\ref{eqn:Ldot}) using the 
parameter values from table \ref{tbl:parms} gives three equilibria: 
(i) a stable \textit{desert state} located at $L^\ast_\text{D}=0$, 
(ii) an intermediate unstable equilibrium at $L^\ast_\text{I}\approx 0.54\,C^\ast_\text{PI}$, 
and (iii) a stable \textit{forest state} at $L^\ast_\text{F}\approx 0.72\,C^\ast_\text{PI}$. 
Hence, our 
carbon cycle 
component
features {\em bistability} between a desirable (forest) and an undesirable (desert) state, 
to one of which the system will 
converge,
depending on initial conditions.

The forest equilibrium 
represents the Holocene carbon cycle
until pre-industrial times, 
neglecting changes in external solar forcing. 
During this period the exchange of carbon between the terrestrial,
maritime,
and atmospheric reservoirs 
were roughly in balance \cite{Ciais2013}. 
The temporal permanence 
during the Holocene is reflected in the model by the 
forest equilibrium's stability. 
The 
model will 
return to the forest state after \textit{small} perturbations which might for instance occur via Volcanic eruptions or other (small) external forcing.

In contrast, the affection of the carbon cycle through human activities like land use (change) and 
GHG emissions 
constitutes a {\em large} perturbation of its natural dynamics. 
To illustrate this, the red arrow depicted in figure \ref{fig:cGLA-phasespace} points from the pre-industrial to the current state, 
far
from the 
forest state and 
already in the basin of attraction of the desert state.

Hence, this simplistic model suggests that the carbon cycle might 
already be in a regime where it would collapse in the {\em future} even without 
further human influence. 
On the other hand, the model does not reproduce well the actual {\em past} evolution of the carbon cycle since the advent of the industrialization, which clearly opposes the shown ``natural'' direction of the flow. For a more reliable analysis, it is thus necessary to explicitly include the human factor into our model, as demanded by \cite{Schellnhuber1998}.

\subsection{How oscillations may emerge in a non-fossil, pre-capitalistic global society}
\label{sec:non-fossil}

We thus 
add
a dynamic 
human population $P$, 
interfering
with the biosphere.
Its only energy source is biomass, no fossil fuels ($a_F=0$) are used yet. 
The global society in this scenario is 
assumed not to accumulate physical capital 
but to operate with a constant amount of 
capital per capita ($K \propto P$). 
Introducing the new parameters $b$ and $y_B$, the expressions for $B$ (\ref{eqn:B}) and $Y$ (\ref{eqn:Y}) read
\begin{eqnarray}
	B_\text{PI} &= b L^\frac{2}{5} P^\frac{3}{5}\;,					\label{eqn:B-PI}\\
	Y_\text{PI} &= y_B B_\text{PI}\;.								\label{eqn:Y-PI}
\end{eqnarray}
In order to reduce the dimension of the model system without altering the qualitative (asymptotic) behaviour, 
the diffusion equilibrium is assumed to establish instantaneously ($d\rightarrow\infty$), 
implying
fixed relations between the carbon stocks, $A=(C^\ast_\text{PI}-L)/(1+m)$ and $M=mA$.
We thus get a two-dimensional system with just $L$ and $P$ as dynamical variables.

\begin{figure*}
	\centering
	\includegraphics[height=0.3\textwidth]{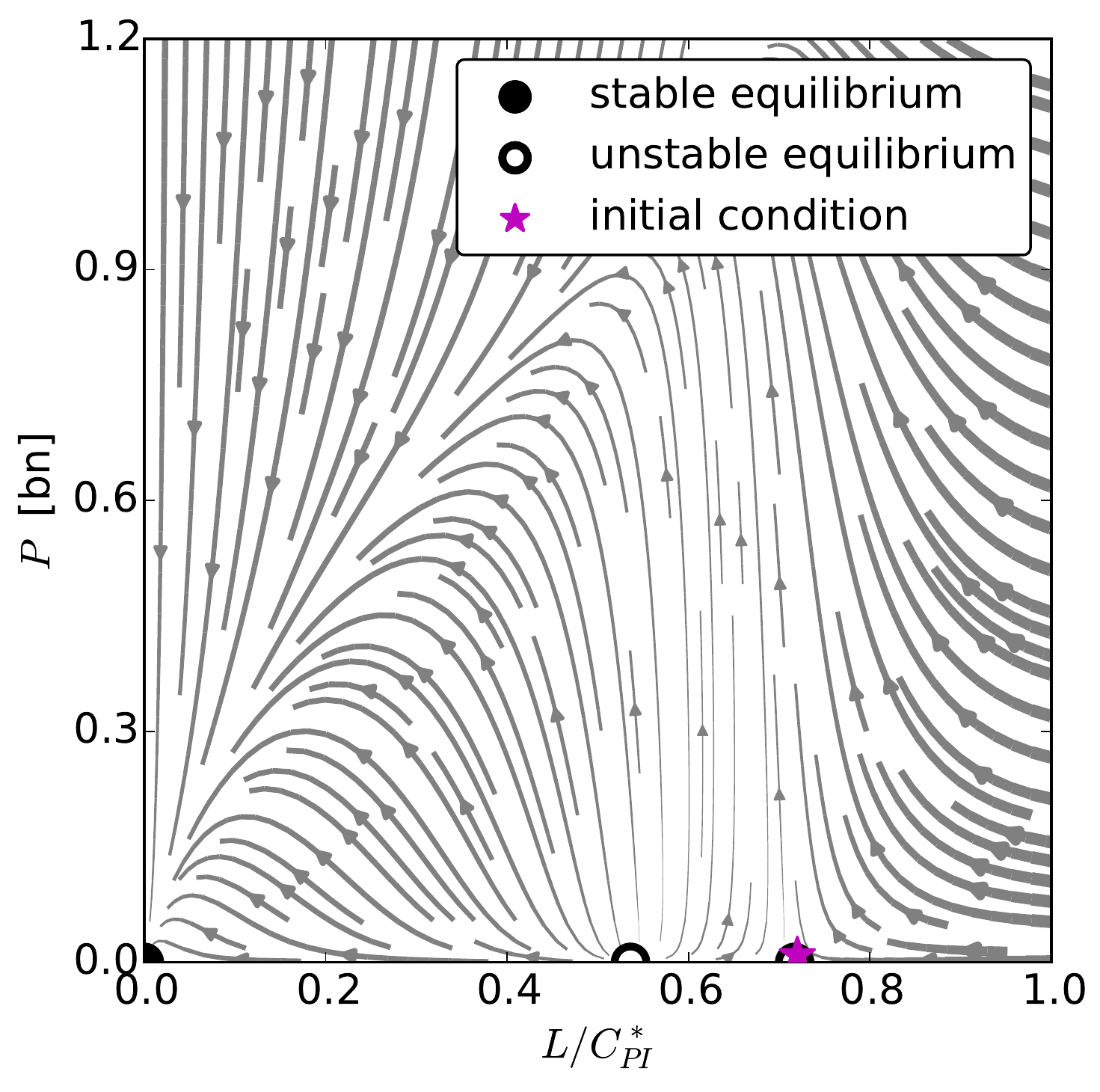}
    \includegraphics[height=0.3\textwidth]{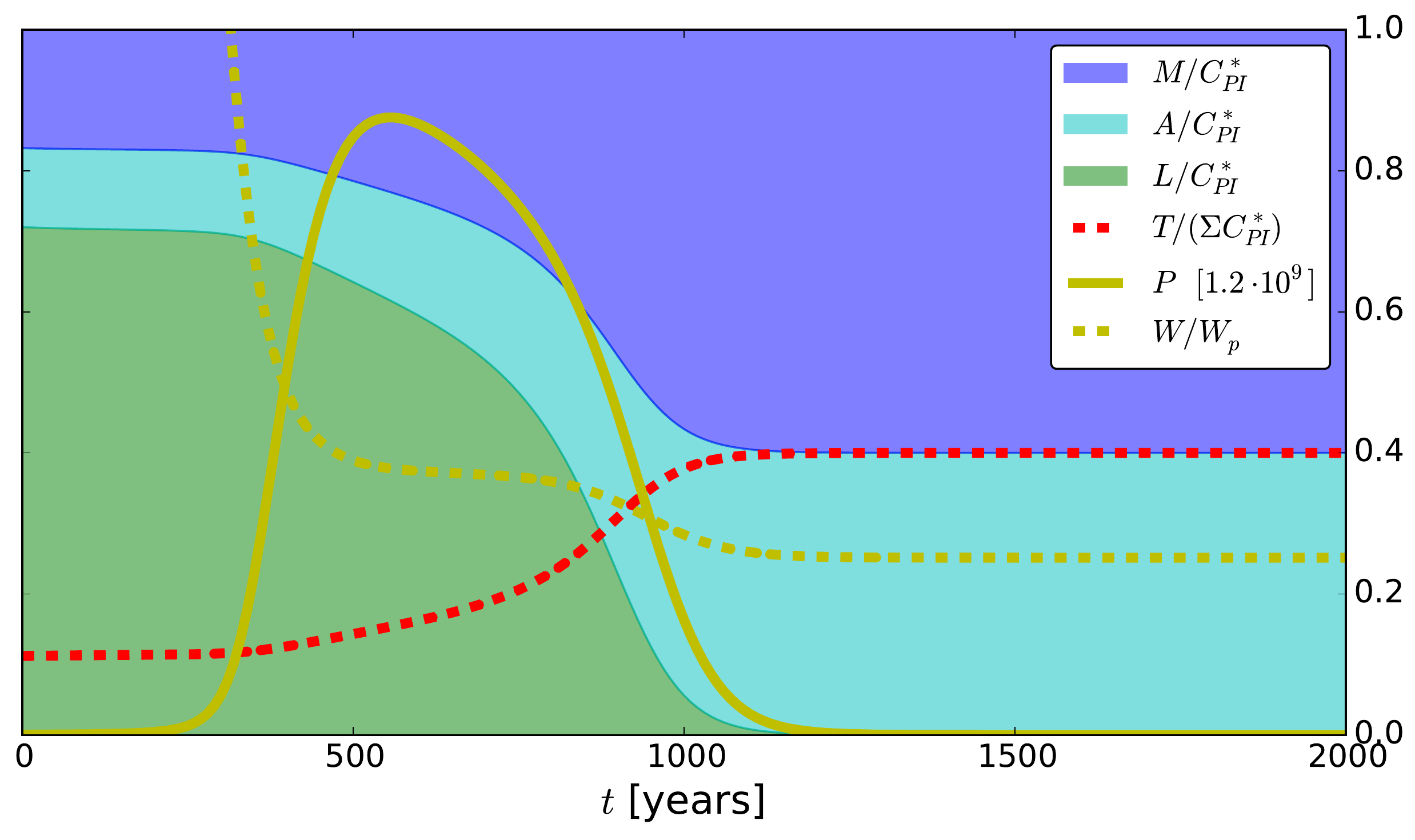}
    \includegraphics[height=0.3\textwidth]{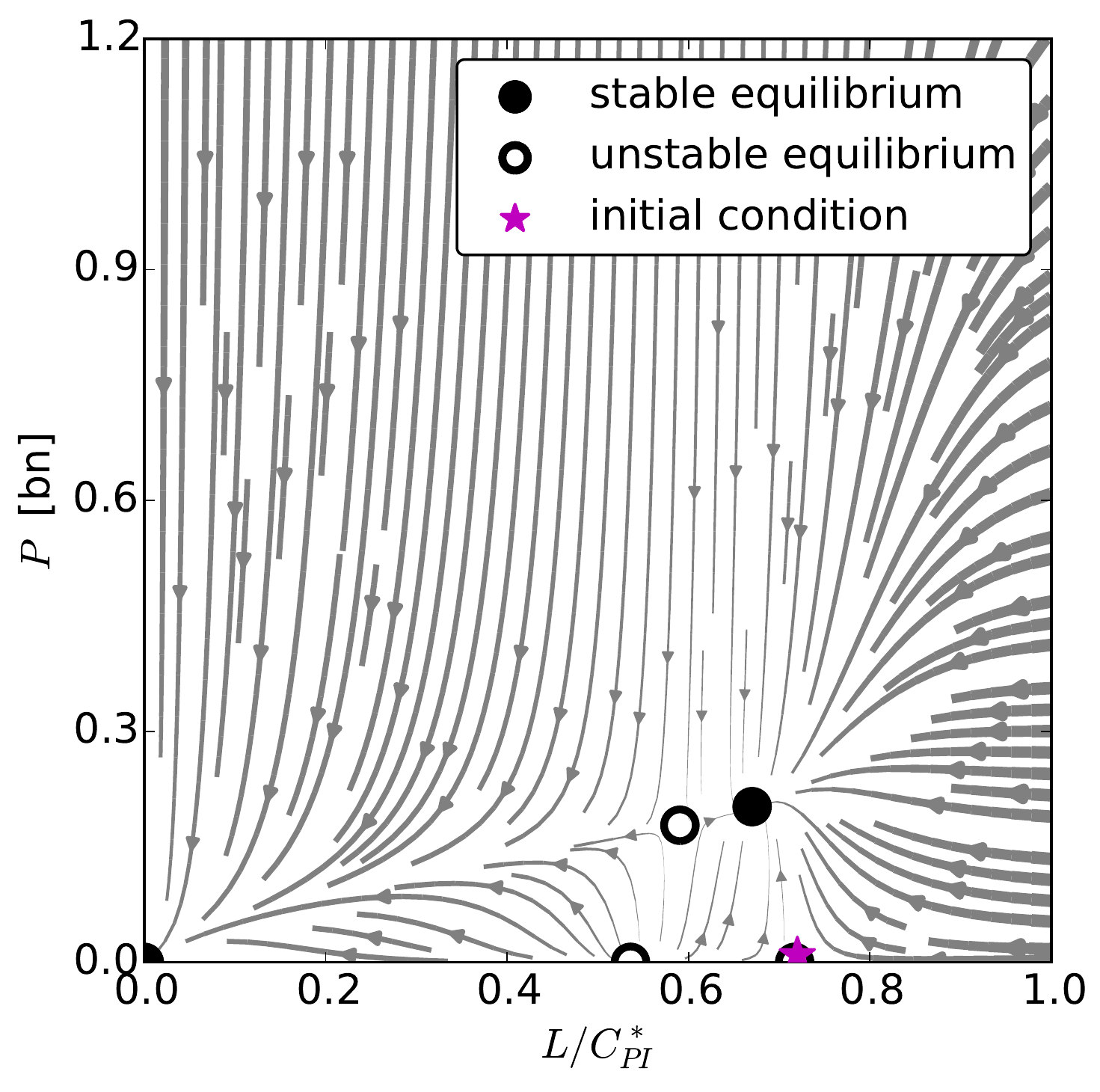}
    \includegraphics[height=0.3\textwidth]{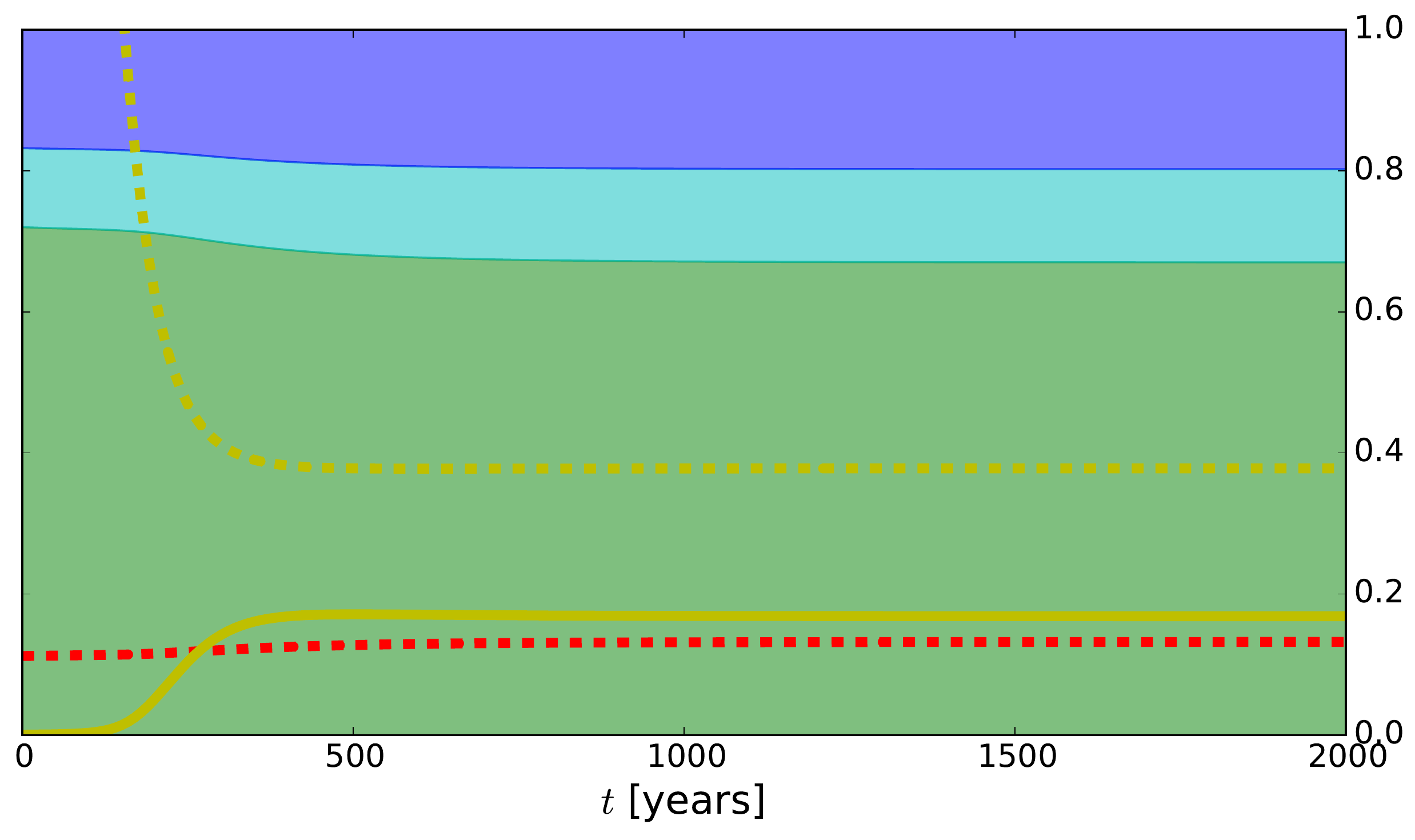}	
	\includegraphics[height=0.3\textwidth]{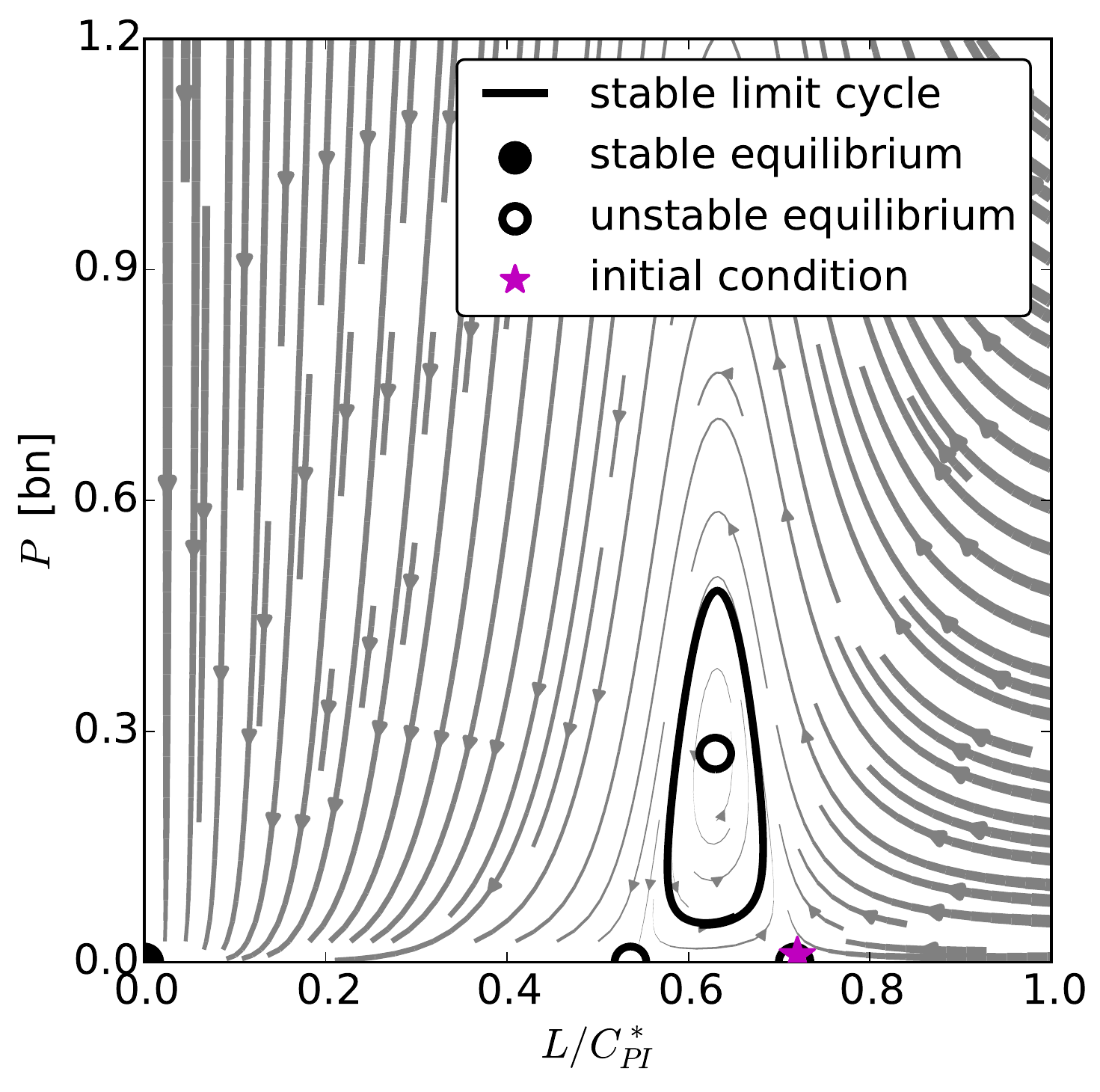}
    \includegraphics[height=0.3\textwidth]{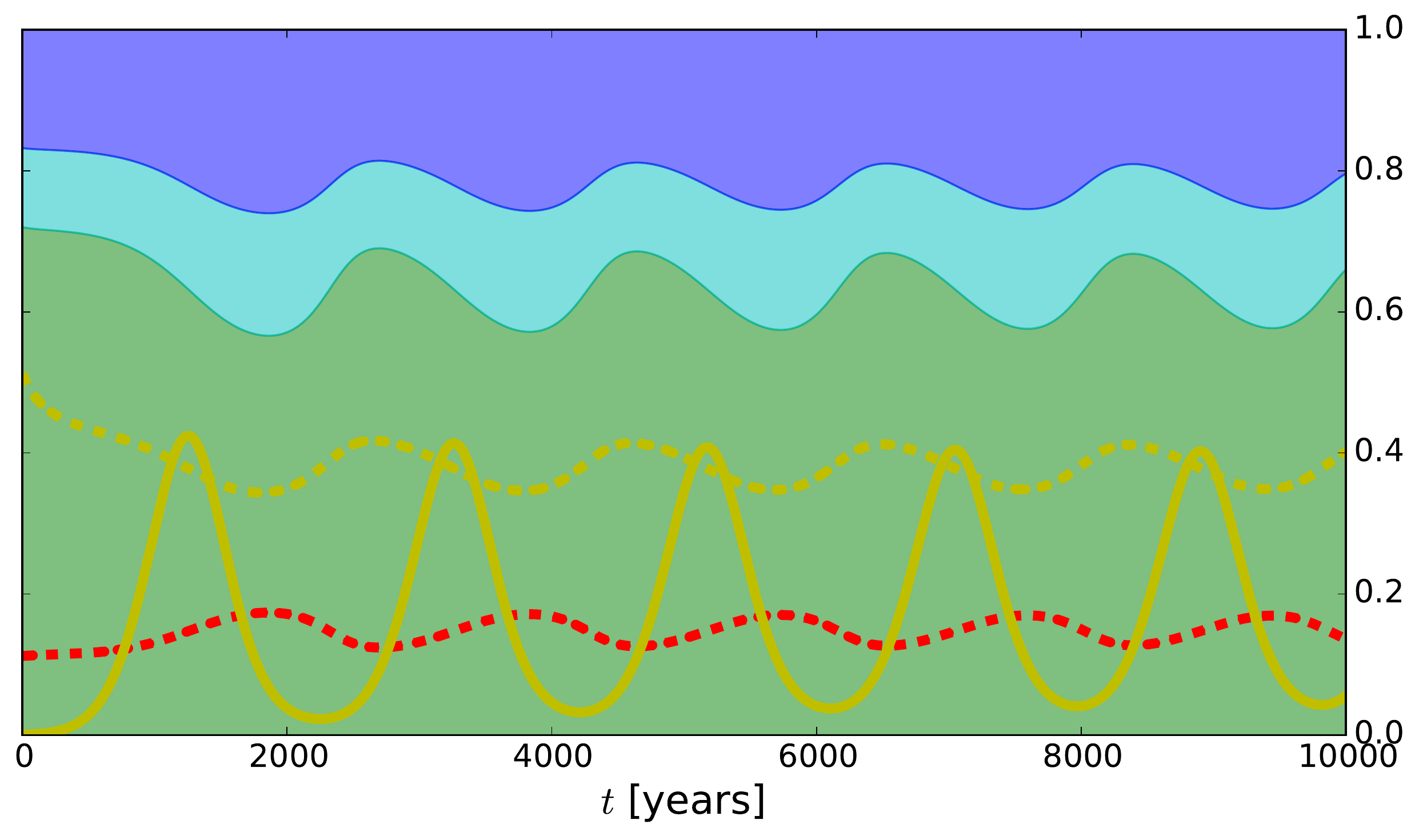}
    \caption{State space representations (left) and exemplary trajectories (right) of the non-fossil, pre-capitalistic model scenario for parameter choices giving rise to qualitatively different asymptotics of the system. In the upper panels the desert state is the only attractor, so that the population overuses natural resources and experiences a global collapse. For a lower economic productivity, shown in the middle panel, the system allows a sustainable coexistence between humans and nature, reflected by the additional attracting equilibrium in the state space. If ecosystem services are considered, an attracting limit cycle can emerge, implying sustained oscillations in all variables with a period of about 2000 years. Note the different scale of the time axis in the lower panel.\newline
    All parameters but the following are set to the default values from table \ref{tbl:parms};
    upper panel: $y_B=2.47\cdot10^{11}\,\$\,\text{GtC}^{-1}$, $w_L=0$;
    middle panel: $y_B=1.235\cdot10^{11}\,\$\,\text{GtC}^{-1}$, $w_L=0$;
    lower panel: $y_B=2.47\cdot10^{9}\,\$\,\text{GtC}^{-1}$, $w_L=4.425\cdot10^{7}\,\$\,\text{km}^2\,\text{GtC}^{-1}\,a^{-1}\,\text{H}^{-1}$.
    Initial conditions: $L_0=2880\,\text{GtC}$, $P_0=500000\,\text{H}$.
    }
    \label{fig:cGLP}
\end{figure*}

In
this pre-industrial scenario one can ask what will ultimately happen to a global society which solely harvests biomass.
The answer
strongly depends on the choice of the parameters. 
Consider 
an initial situation 
with 
$P_0=500000$ 
on
a forested planet ($L_0=0.72\,C^\ast_\text{PI}=2880\,\text{GtC}$); 
furthermore all parameters are set to the default values (cf. table \ref{tbl:parms}) and ecosystem services are neglected ($w_L=0$) (figure \ref{fig:cGLP}, upper right panel). 
Due to the abundance of resources, 
the population initially prospers and grows (exponentially) fast. 
Biomass use also increases but slower than population (equation (\ref{eqn:B-PI})), so that well-being decreases as a consequence (equation \ref{eqn:W}); 
this in turn lets the population growth rate decrease. 
After about 600 years a maximum population of about one billion humans is reached while the terrestrial carbon stock is considerably lower than initially. 
Despite the following decrease in population, the pressure on the 
ecosphere by humans
pushes the carbon cycle into the basin of attraction of the (undesirable) desert state and an unpopulated planet prevails after about 1200 years.
When regarding the 
state space of the system (figure \ref{fig:cGLP}, upper left panel) it becomes clear why this \textit{collapse} was inevitable. 
There simply is no \textit{coexistence equilibrium} 
with
$L>0$ and $P>0$, 
and even the two unpopulated forest equilibria with $L>0$ and $P=0$ are unstable (one in the $L$-, the other in the $P$-direction) so that only the desert state equilibrium at $L=P=0$ is 
an
attractor. 
Hence independent of the initial conditions the system will ultimately evolve to the desert state.

While such collapse has been observed historically for 
\textit{local} agricultural civilizations \cite{Brander1998}, 
a global collapse of the terrestrial ecosystems did not occur so far. 
For slightly altered parameter values, an evolution of the model system occurs which matches the historic one better, until the onset of the industrialization. 
However,
if the value of $y_B$ (whose estimate 
has
a high uncertainty) is halved, 
a \textit{sustained coexistence} between the terrestrial ecosystems and the human population becomes possible (figure \ref{fig:cGLP} middle panels). 
In addition to the three equilibria at the $P$-axis ($P=0$), there exist two equilibria with $L>0$ and $P>0$ of which one is stable. 
Starting from the same initial state as above the system initially behaves similar, 
but the population rise is less extreme and humans 
exert less pressure on the terrestrial carbon stock.
After about 400 years an equilibrium with constant carbon stocks, population and well-being
is reached. 
The asymptotic population 
of about 200 mn 
compares nicely with actual estimates of the global population in medieval times \cite{Kremer1993}, 
for which the non-fossil, pre-capitalistic model scenario seems adequate. 
A long period of stagnating socio-economic observables is also in line with the Malthusian population model \cite{Malthus1798}.

Like Malthus,
we identified well-being (which determines fertility and mortality, cf. (\ref{eqn:Pdot})) with per-capita consumption
so far. 
It is, however, reasonable to assume that the integrity of nature also contributes to human well-being via 
ecosystem services \cite{MEA2005,Haines-Young2010}.
Hence we 
consider a third setting in which well-being is dominated by ecosystem services by choosing $w_L>0$ and a low value for $y_B$ (figure \ref{fig:cGLP}, lower panels). 
The phase portrait qualitatively differs from both previous cases as it features an attracting \textit{limit-cycle} but no stable coexistence equilibrium. 
Hence there are trajectories -- such as the shown one -- which are characterized by sustained \textit{oscillations} of all variables. 
As before,
population rises 
until it reaches a maximum of about 500 mn humans after about 1500 years. 
The growing biomass consumption is accompanied by decreasing well-being and -- with a short delay -- decreasing population. 
$P$ declines until it reaches a minimum after another approximately 800 years, 
now 
taking pressure from the terrestrial carbon stock, which is thus able to recover. 
This in turn directly increases well-being via the contribution of ecosystem services, allowing 
population to recover
as well. 
These feedbacks lead to 
oscillations with a period of about 2000 years. 
Qualitatively, the observed patterns are very similar to those described in classical models of predator-prey ecosystems \cite{Lotka1910}.
In contrast to the latter, however, our model is still multistable in this regime
since the ``desert'' equilibrium is still also stable due to the functional forms for fertility and economic production. 
Other models of human-nature coevolution feature oscillations \cite{Brander1998, Kellie-Smith2011, Motesharrei2014}
which may be sustained or dampened but typically have shorter periods. 

The presented parameter settings and trajectories are of course just exemplary and hence their quantitative implications should not be overrated. 
There
are also intermediate cases for which dampened oscillations occur, 
not shown 
here
since the asymptotic states are unchanged.

\begin{figure}
	\centering
    \includegraphics[width=0.48\textwidth]{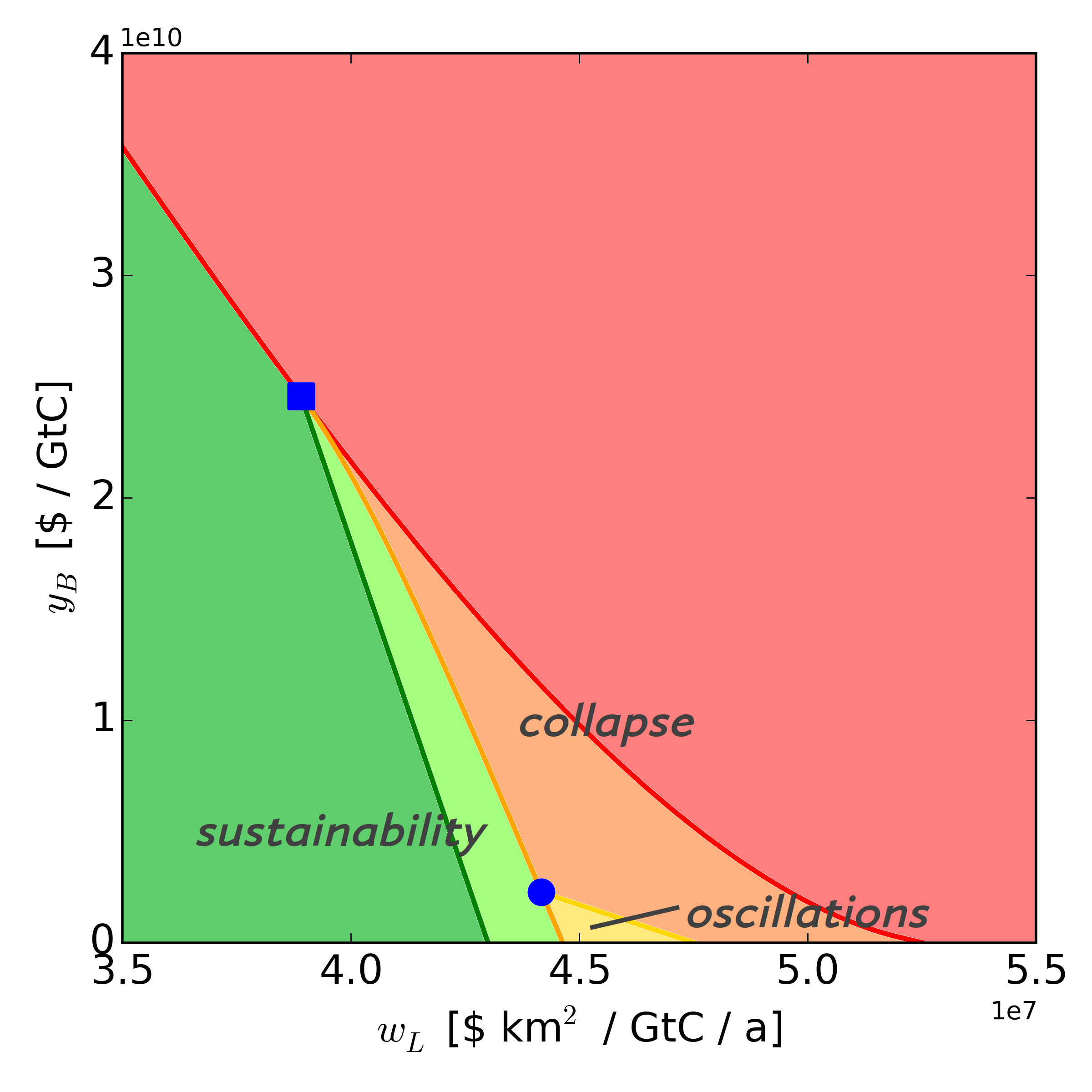}
    \caption{Bifurcation diagram in the $(y_B,w_L)$ parameter space, showing five qualitatively different dynamic regimes. While within the two greenish regimes a \textit{sustainable} (stable) coexistence of nature and society is possible for some initial conditions, both will \textit{collapse} in the two reddish parameter regimes. Only the very small regime indicated in yellow features sustained \textit{oscillations} in the dynamic variables. Borders between the regimes correspond to different local or global bifurcation curves. For details and differences within the greenish and reddish regimes, cf.\ \ref{sec:bifurcations}. 
    }
    \label{fig:bifurcation-diagram}
\end{figure}

The qualitative changes of the asymptotic behaviour of the system under variation of parameters can be analysed mathematically using 
bifurcation theory \cite{Kuznetsov1998}. 
A more rigorous study reveals that there are indeed five different regimes in the $(y_B,w_L)$ parameter space,
with
qualitatively different asymptotic states. However, there are only three different regimes (sustainability, collapse, oscillations) for which there are different {\em attracting} asymptotic states,
as
discussed above.
The 
bifurcation diagram is shown in figure \ref{fig:bifurcation-diagram},
the full bifurcation analysis is in \ref{sec:bifurcations}.

\subsection{Possible collapse of a fossil-based, capitalistic global society}
\label{sec:fossil}

We finally consider a scenario which extends the 
previous
one in two ways.
First, in addition to biomass use ($B$) now also fossil fuel extraction ($F$) from the geological pool $G$ is enabled, where the relative shares of the two energy sources is determined by a price equilibrium.
Second, physical capital $K$ is now 
a stock variable with a standard growth dynamics decoupled from population growth. 
Altogether, this scenario applies to the era since the onset of the industrialization until recent times
during which biomass and fossil fuels are the dominant energy sources and physical capital became a major factor of production. 
Moreover, we drop the assumption of the diffusion equilibrium from the previous scenario, giving a less stylized and more realistic representation of the global carbon cycle. 
Thus we have the full five-dimensional dynamical system ($L$, $A$, $G$, $P$, $K$) given by (\ref{eqn:Ldot}) to (\ref{eqn:Kdot}).

\begin{figure*}
	\centering
    \includegraphics[height=0.3\textwidth]{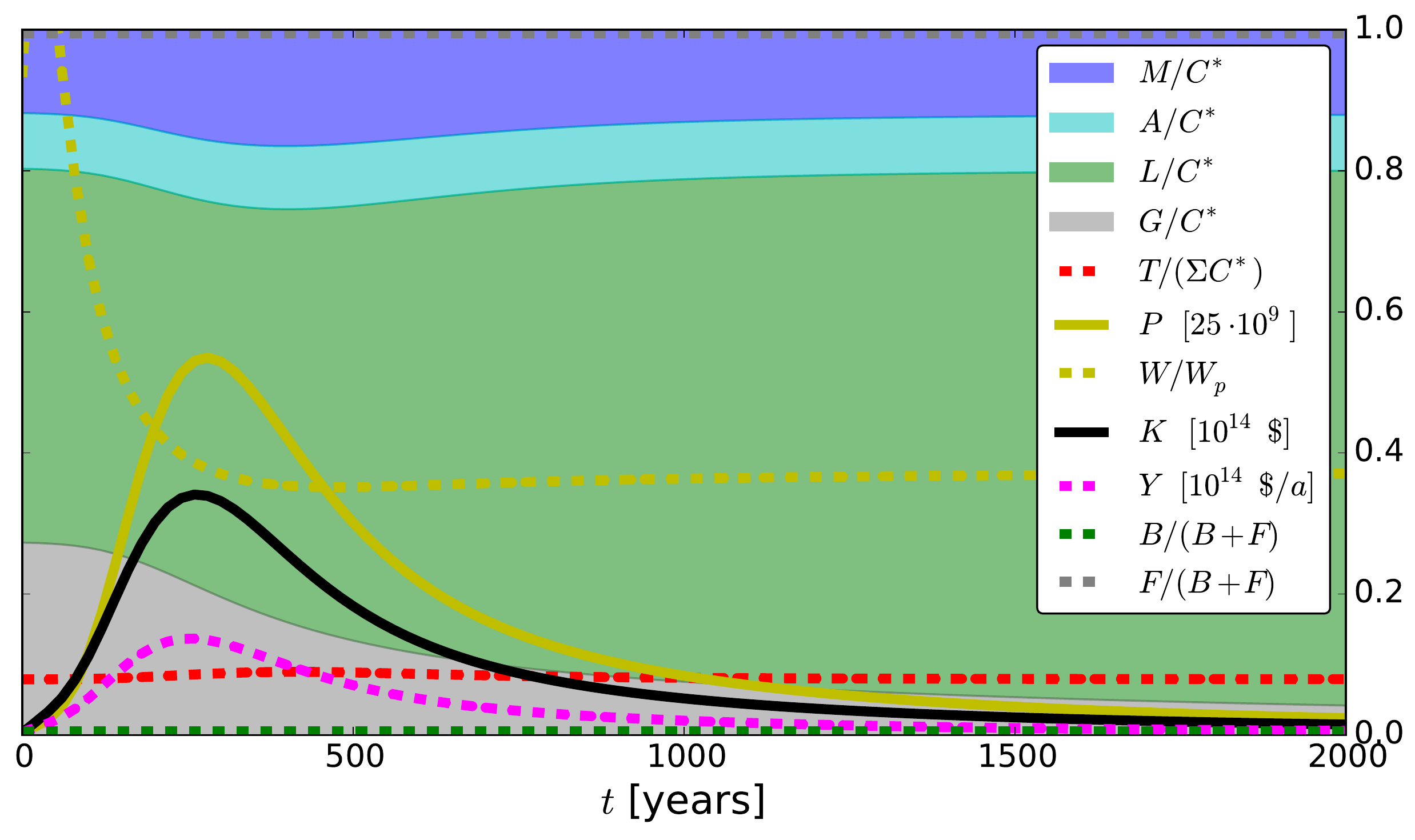}\\
    \includegraphics[height=0.3\textwidth]{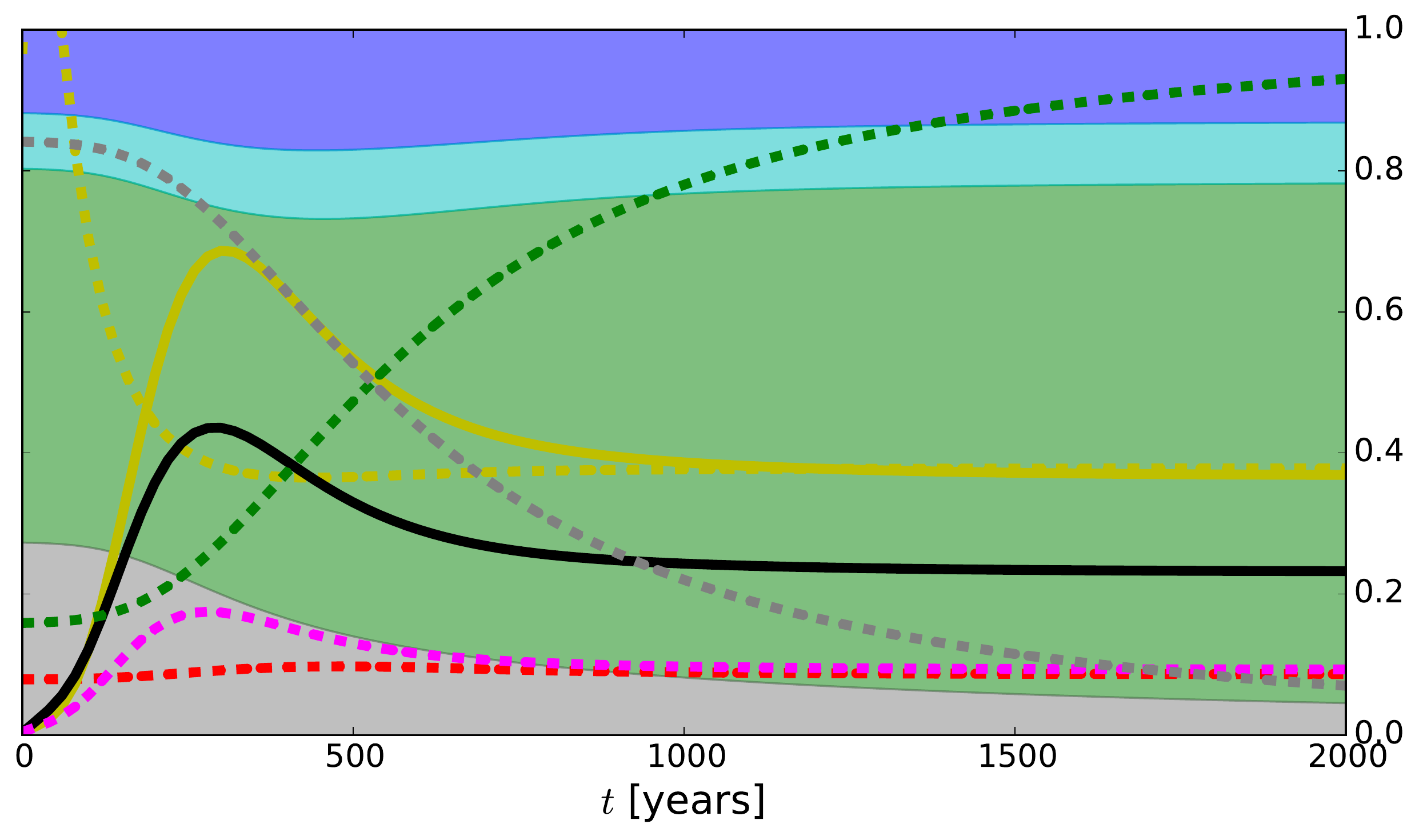}\\
    \includegraphics[height=0.3\textwidth]{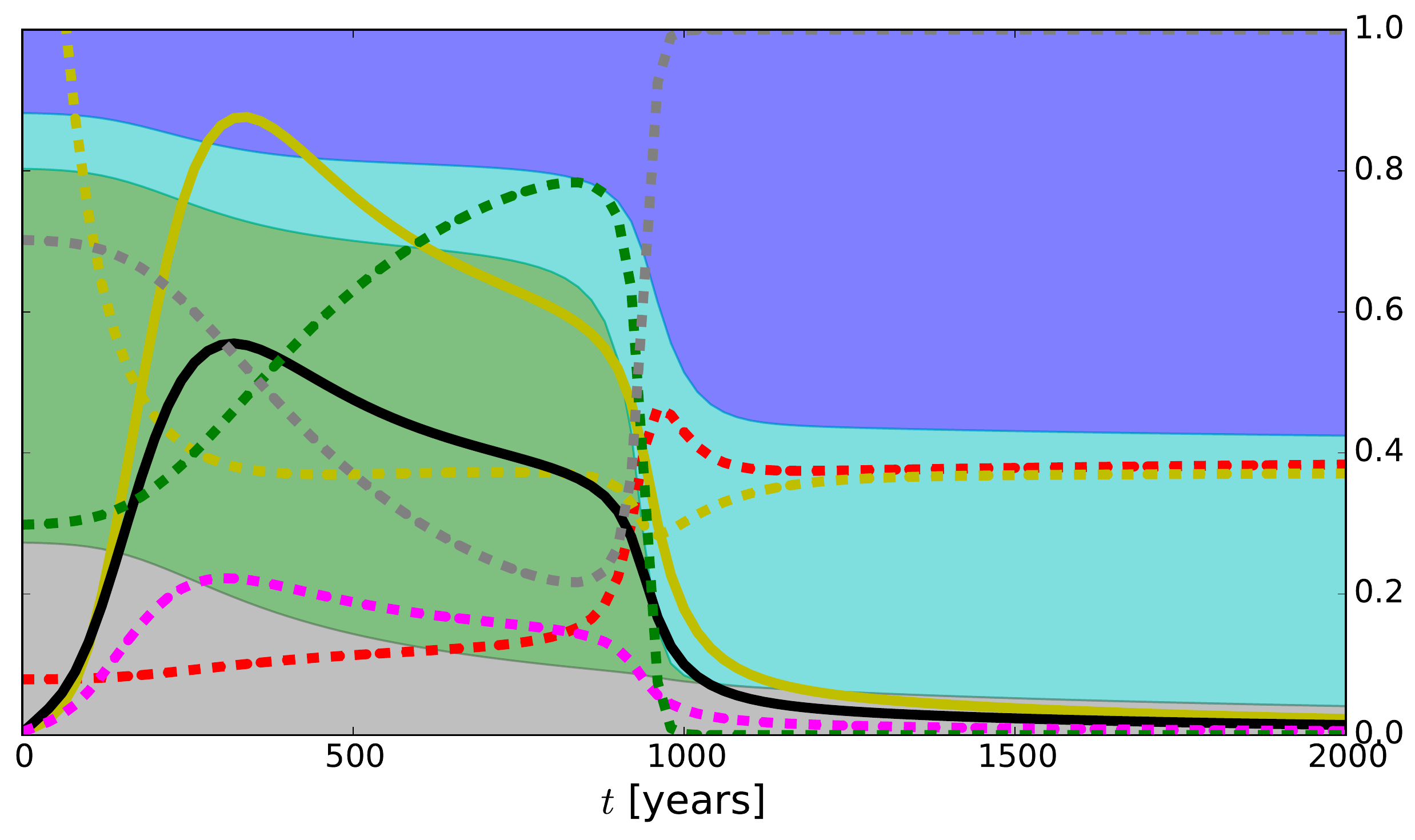}
    \caption{Exemplary trajectories of the fossil-based, capitalistic model scenario for different usage of biomass and fossil fuels reflected by different combinations of the sector productivities $a_B$ and $a_F$. In the fossil-only reference setting (upper panel) the global will go extinct after several millennia with the depletion of the geological carbon stock while the emitted carbon is mainly stored in the terrestrial stock. Moderate usage of biomass allows a sustained coexistence of humans and nature in the long run (middle panel) but fossil resources will still completely depleted. When humans exert too much pressure on the terrestrial system through biomass use (land use) these can ultimately collapse, thereby ruining the preconditions for life on Earth (lower panel). The socio-economic development is indistinguishable in the scenarios with enabled biomass use until about 800 years of simulation time. Only changing the continued changes in the natural subsystem of Earth indicate the prolonged transient towards an undesirable desert state.\newline
    All parameters but the following are set to the default values from table \ref{tbl:parms};
    upper panel: $a_F=24.9\,\text{GJ}^5\,\text{a}^{-5}\,\text{GtC}^{-2}\,\$^{-2}\,\text{H}^{-2}$, $a_B=0$;
    middle panel: $a_F=24.9\,\text{GJ}^5\,\text{a}^{-5}\,\text{GtC}^{-2}\,\$^{-2}\,\text{H}^{-2}$, $a_B=1.25\,\text{GJ}^5\,\text{a}^{-5}\,\text{GtC}^{-2}\,\$^{-2}\,\text{H}^{-2}$;
    lower panel: $a_F=24.9\,\text{GJ}^5\,\text{a}^{-5}\,\text{GtC}^{-2}\,\$^{-2}\,\text{H}^{-2}$, $a_B=2.8\,\text{GJ}^5\,\text{a}^{-5}\,\text{GtC}^{-2}\,\$^{-2}\,\text{H}^{-2}$.
    Initial conditions: $L_0=2915\,\text{GtC}$, $P_0=162\cdot10^6\,\text{H}$, $K_0=323\cdot10^9\,\$$
    }
    \label{fig:cGLGPK}
\end{figure*}

The availability of two different energy forms gives rise to the following question which connects closely to the introductory question of the previous section: 
What is the ultimate fate of the human population for different usage patterns of biomass and fossil fuels?

The proneness to use a certain form of energy is determined by various factors (cf. (\ref{eqn:B}), (\ref{eqn:F})).
It increases with the size of the associated stock variable ($L$ for biomass, $G$ for fossil fuels) and with the respective productivities ($a_B$, $a_F$),
but decreases because of substitution effects the cheaper the other energy form is.
While the stock sizes $L$ and $G$ are prescribed by the natural Earth system, $a_B$ and $a_F$ are rather abstract economical parameters which are hard to estimate from real-world data. 
The choice of their absolute and relative values hence facilitates an investigation of different energy usage scenarios. 
In the subsequent analysis the effects introduced by regarding ecosystem services discussed in section \ref{sec:non-fossil} are excluded by setting $w_L=0$.

To isolate the effect of emissions caused by 
fossil fuels, we 
regard
a reference setting in which biomass use is disabled ($a_B=0$) and the fossil fuel sector productivity is set to a value for which the extraction speed of fossils roughly coincides with observed values over the past 250 years ($a_F=24.9\,\text{GJ}^5\,\text{a}^{-5}\,\text{GtC}^{-2}\,\$^{-2}\,\text{H}^{-2}$).
The abundance of resources causes population and physical capital to grow fast initially until they reach a maximum after about 300 years (figure \ref{fig:cGLGPK}, upper panel). After this initial boom, well-being 
saturates, then
both $P$ and $K$ slowly decrease and the economic production $Y$ is reduced accordingly. This slow perishing of the economy and population is due to the dependence on fossil fuels from the 
non-renewable
geological carbon stock $G$.
After 
2000 years 
the population is close to extinction and fossil fuels are almost 
depleted. Notably, for this choice of parameters, the emissions of fossil carbon only lead to a slight increase of the atmospheric carbon content (and the associated global mean temperature), while most of the carbon is captured in biomass and soils. Also for other values of $a_F$, a collapse of the terrestrial system to a desert state due to emissions of fossil fuels is not observable in the model. 
However, the fate of a population in this purely fossil-based scenario is slow extinction on a well-forested planet, but now with an almost unchanged level of well-being until the end.

Obviously, this scenario is not very realistic since humans would certainly start to (and historically always did) harvest biomass in order to satisfy their need for energy. By choosing a rather low biomass sector productivity of $a_B=0.05\,a_F$ the initial share of biomass in total energy use amount to about $15\,\%$ (figure \ref{fig:cGLGPK}, middle panel). 
The behaviour of the system during the first 500 years of simulation time is very similar to the reference setting with the only difference that, due to the additional use of biomass, $P$, $K$ and thus $Y$ reach higher absolute levels.
Due to the depletion of the geological carbon stock and the increase in terrestrial carbon, the share of biomass is constantly increasing and overtakes the fossil share after about 500 years. In contrast to the previous setting the global society has an alternative to fossil fuels and is not doomed to go extinct. Instead, the population decrease slows down and a sustained coexistence between humans and nature emerges. Note that humans still continue to use fossil fuels until ultimately the geological carbon stock is completely depleted, which follows from the economical model of the energy sector ($F=0$ necessitates $G=0$ as long as $P,K>0$, cf. (\ref{eqn:F})). An abandoning of fossil fuel use can thus not be achieved by the economic forces assumed in the model; instead this would necessitate  other economical mechanisms, e.g., banning or taxing of fossils through policies. In the asymptotic state about 10 bn humans inhabit Earth, the average per-capita capital amounts to about $2500\,\$$ which we regard as realistic orders of magnitude. 



So can we conclude that biomass use can save humankind when fossils are abandoned for whatever reason? This must clearly be denied as our last parameter setting shows, in which assume a larger biomass sector productivity ($a_B=0.1125\,a_F$). 
Now biomass initially makes up about a third of the total energy used and becomes the dominant form of energy after about 350 years. Again the socio-economic observables ($P$, $K$, $Y$) behave qualitatively very similar to the previous settings (fast increase to a maximum, followed by slowing decrease) until about 800 years of simulation time. About this time their speed of decrease accelerates again and they drop to very low values within about 200 years. This breakdown of the socio-economic system is caused by overuse of natural resources which triggered a collapse of the biosphere (represented by the terrestrial carbon stock $L$) to the desert state, just as observed in the non-fossil, pre-capitalistic scenario discussed in section \ref{sec:non-fossil}.
After the collapse humans can only ``survive'' until the remaining fossil fuel resources are completely depleted, 
so that ultimately, an unpopulated desert planet prevails.
This is, of course, not realistic for several reasons:
the life-enabling capacity of the biosphere (e.g., through oxygen production) is not accounted for and renewable energy is not available in the model.
We thus learn that the intensity of biomass and land use, reflected by the parameter $a_B$ are of crucial importance for a sustainable global coevolution of humans and nature which should always be considered besides the necessity for reducing emissions from fossil fuels. 
While the parameter value is fixed in the model simulation, in reality, the socio-economic conditions it reflects can be subject to change, e.g., through policy instruments. 

It should be pointed out that the collapse of the system in the third setting could not have been predicted by looking solely at socio-economic observables, 
as these evolve analogously in the previous settings for roughly the first 800 years of simulation time. 
Merely the changing environmental conditions, as indicated by the continued increase in global mean temperature and decrease of the vegetation
from year 300 to 800, qualitatively differentiate this setting from the previous ones and thus hint at the fact that the system actually undergoes a long transient period towards an undesirable final state.
Note that we do not even need to model direct climate damages on, say, mortality and capital depreciation, to cause the extinction.

A second question posed by the industrialization scenario is: What is the effect of the dynamic physical capital stock $K$, compared to the non-capitalistic societies discussed above?
For all regarded parameter settings population and capital evolve alike, meaning a constant capital per capita just as it was assumed in the previous, non-capitalistic scenario. This observation can be explained with the rate of capital depreciation ($k$) which is comparable to the reproduction rates of humans. A considerably lower depreciation rate would instead introduce a time lag between the trajectories of $P$ and $K$. The estimated parameters, however, indicate rather short time scales for the changes of the factors of production, compared to the rather slow evolution of the carbon stocks (apart from collapses).

\section{Conclusions}
\label{sec:conclusions}

We presented a flexible conceptual World-Earth model which is -- through an appropriate choice of variables and parameters -- able to qualitatively represent the global coevolutionary dynamics of humans and nature for different socio-cultural stages of human history on Earth, particularly during the Holocence and Anthropocene epochs. 
The actual evolution of global carbon stocks was found to oppose the dynamics to be expected from the topology of the natural carbon cycle, which is mainly due to human interference with natural dynamics through land use (change) and emissions of carbon into the atmosphere. Due to various nonlinearities in natural and social dynamics, an accurate description of the mid and long-term evolution of the Earth system thus necessitates an explicit modelling of the ``human factor'' with a balanced representation of natural and socio-economic subsystems. Our conceptual model (framework) thus contributes to the challenge of ``Modelling the Anthropocene''.

For each model scenario we identified the characteristics of possible asymptotic states of the system which comprise a sustainable coexistence of humans and nature, a collapse of both natural and socio-economic subsystems and even persistent oscillatory dynamics with multi-millennial periods. By systematic variation of those parameters whose estimates from real-world data are particularly uncertain, we found the preconditions of the different asymptotic patterns. It is especially those parameters related to the appraisal ($w_L$) or the intensity of use ($y_B$, $a_B$) of the biosphere, which make a crucial difference for the fate of the planet and humankind.

The overall picture of our results supports the insight that neither fossil fuels nor biomass use are likely to facilitate a sustainable coexistence of several billion humans on a planet with limited natural resources. We conclude that besides reducing the global demand for energy, merely the extensive use of renewable energy forms may pave the way into a sustainable future of a well-developed global society. Extending the current framework by enabling the use of renewables is thus a priority for the future model development.

In our model analysis we focussed mainly on understanding the asymptotic behaviour of the coevolutionary Earth system and hence regarded rather long time scales of several centuries to millennia. A lot of interesting dynamics like growth phases or collapses, can, however, happen on quite short time scales from decades to centuries. These transient phases could reveal interesting insights, particularly regarding the evolution of the socio-economic subsystem of the Earth. We believe that historically observed phenomena like the ``Great Acceleration'' \cite{Steffen2015a} could, in principle, be reproduced with our model, given appropriate parameter values and initial conditions. To show this, in \ref{sec:superexponential} we derive conditions under which the socio-economic observables of the model ($K$, $Y$, $P$) feature super-exponentially fast growth.

Beyond the implications for global sustainability our simple models emphasize the subtleties resulting from the nonlinear characteristics of the Earth system, e.g., depicted by very long-lasting transients towards undesirable attractors. Realizing that such dynamical features can even emerge in simple conceptual models like the presented ones, should raise the awareness and caution also for the analysis of more comprehensive models of the Earth system.

\clearpage
\ack

We acknowledge support by the German Research Foundation and the Open Access Publication Funds of the G\"ottingen University.

We further thank Wolfram Barfuss, Jonathan Donges, Vera Heck, Tim Kittel and the whole COPAN group at PIK for helpful discussions and comments throughout all stages of this work.

\section*{References}
\bibliographystyle{my-iopart-num}
\bibliography{references}

\clearpage
\appendix

\section{Derivation of the model}
\label{sec:model-derivation}

\def\ge{\geqslant}
\def\geq{\geqslant}
\def\le{\leqslant}
\def\leq{\leqslant}

\subsubsection*{Variables.}

The two main variables of interest for this model are 
human well-being $W$ (representing the most important aspect of the anthroposphere or socio-economic subsystem of Earth)
and terrestrial carbon stock $L$ (representing the most important aspect of the ecosphere or biophysical subsystem of Earth). 
We try to restrict the model to those further variables and processes that seem indispensable
in order to assess the qualitative features of the possible coevolutionary pathways of $L$ and $W$
on a time-scale of hundreds to thousands of years, hence we include the following quantities needed to represent 
a carbon cycle and resource-dependent economic and population growth:
\begin{itemize}
  \item Time $t$ [standard unit: years, a]
  \item Terrestrial (``land'') carbon stock $L\in [0,C^\ast]$ [GtC] (including soil and plants)
  \item Atmospheric carbon stock $A\in [0,C^\ast]$ [gigatons carbon, GtC]
  \item Accessible geological carbon stock (serving as fossil fuel reserves) $G\in [0,C^\ast]$ [GtC]
  \item Maritime carbon stock $M = C^\ast - A - G - L\in [0,C^\ast]$ 
        [GtC] (including only the upper part of the oceans which exchanges carbon comparatively fast with air)
  \item Human population stock $P\ge 0$ [number of humans, H]
  \item Physical capital stock $K\ge 0$ in time-independent (e.g. 1992) US dollars
  \item Global mean surface air temperature $T\ge 0$ (representing ``climate''), 
        measured not in Kelvin but for simplicity in ``carbon-equivalent degrees'' [Ced$=$GtC],
        using an atmospheric carbon-equivalent scale. 
        I.e., $T=x\,$Ced is the equilibrium temperature of an atmosphere containing $x\,$GtC).
  \item Biomass extraction flow $ B\ge 0$ [GtC/a] and biomass energy flow $ E_B\ge 0$ [GJ/a]
  \item Fossil carbon extraction flow $ F\ge 0$ [GtC/a] and fossil energy flow $ E_F\ge 0$ [GJ/a]
  \item Total energy input flow $ E\ge 0$ [GJ/a]
  \item Economic output flow $ Y\ge 0$ [\$/a]
  \item Investment flow $ I\ge 0$ [\$/a]
  \item Well-being $ W$ in per-capita consumption-equivalent units 
        [\$/a\,H] (including economic welfare and environmental effects, e.g., health and ecosystem services)
\end{itemize}
We follow the predominant economic convention of measuring capital, production, and consumption in monetary units.
$ A, B, E, F, G, I, K, L, M, P, Y$ are {\em extensive} quantities 
in the sense that the would double if the Earth System was replaced by two identical copies of itself,
while $ T$ and $ W$ are {\em intensive} quantities which would not double.
The only {\em conserved} quantity in the model is carbon, as expressed by the equation $ A +  G +  L +  M \equiv  C^\ast$.

\subsubsection*{Processes, generic interaction terms and equations.}

The following processes and dependencies are considered to be the main drivers of the carbon cycle, economic and population growth:
\begin{itemize}
  \item Ocean to air diffusion $ f_\text{diff.}( A, M)$ [GtC/a] (ignoring pressure and temperature dependency)
  \item Greenhouse effect on temperature $T=T( A)$ [GtC] (ignoring other GHG)
  \item Land to air respiration $ f_\text{resp.}(L, T)\ge 0$ [GtC/a] (ignoring other dependencies)
  \item Photosynthesis $ f_\text{photos.}( A, L, T)\ge 0$ [GtC/a] (ignoring nitrogen and other dependencies)
  \item Biomass extraction $ B =  B( G, K, L, P)\ge 0$
  		and combustion $ E_B =  E_B( B)$ (ignoring other economic dependencies, 
  		and afforestation, carbon storage and other policy dependencies, 
  		and assuming almost all extracted land carbon ends up in the atmosphere after a negligible time;
  		ignoring carbon stored in human bodies and physical capital)
  \item Fossil fuel extraction $ F =  F( G, K, L, P)\ge 0$ 
        and combustion $ E_F =  E_F( G)$
  \item Total energy usage from these energy sources $ E =  E_B +  E_F$.
  \item Economic production of output $ Y =  Y( E, K, P)$ (assuming the two energy sources are perfect substitutes)
  \item Capital growth through investment $ I = i  Y$
  \item Capital depreciation $f_\text{deprec.}( K)\ge 0$ [\$/a]
  \item Consumption of all non-invested economic output and emergence of well-being $ W =  W( L, P, Y)$
  \item Population fertility and mortality $ f_\text{fert./mort.}( W)$ [1/a]
\end{itemize}
This leads to the following generic equations:
\begin{eqnarray}
    d L/d t &=  f_\text{photos.}( A, L, T) -  f_\text{resp.}( L, T) -  B,\\
    d A/d t &=  -dL/dt +  F +  f_\text{diff.}( A, M),\\
    d G/d t &= -  F,\\
    d K/d t &= i  Y -  f_\text{deprec.}( K),\\
    d P/d t &= ( f_\text{fert.}- f_\text{mort.})( W) P
\end{eqnarray}
with
\begin{eqnarray}
     T &= T( A),\\
	 B &=  B( G, K, L, P),\\
	 F &=  F( G, K, L, P),\\
	 E &=  E_B( B) +  E_F( F)\\
	 Y &=  Y( E, K, P),\\
	 W &=  W( L, P, Y).
\end{eqnarray}

\subsubsection*{Choice of functional forms.}

Since our aim is a mainly qualitative analysis rather than quantitative prediction,
we aim at choosing simple functional forms that fulfil at least the following qualitative properties:
\begin{itemize}
  \item $ f_\text{diff.}$ is increasing in $ M$ and decreasing in $ A$.
  \item $ T$ is increasing in $ A$.
  \item $ f_\text{resp.}$ is roughly proportional to $ L$ and is increasing but concave in $ T$ 
  			(over the range of temperatures experienced in the holocene).
  \item $ f_\text{photos.}$ is roughly proportional to $ L$,
  			is increasing and concave in $ A$ (due to diminishing marginal crabon fertilization), 
            and is decreasing in $ T$ (over the range of temperatures experienced in the holocene).
  \item $ f_\text{deprec.}$ is roughly proportional in $ K$.
  \item $ f_\text{fert.}$ is zero for vanishing $W$, 
  			grows roughly proportionally with $W$ for small values of $W$ 
  			(representing basic nutritional needs for reproduction as in ecological models),
  			grows more concavely when $W$ grows further until $W$ reaches some value $W_P>0$
  			(representing saturation of fertility due to biological limits)
  			and finally declines again towards zero when $W$ grows even further
			(due to education- and social security-related effects).	
  \item $ f_\text{mort.}$ is infinite for vanishing $W$ and declines towards zero with growing $W$.
  \item $ E_B, E_F\ge 0$ are roughly proportional to $ B$ or $ F$, respectively.
  \item $ B$ is increasing in $ K, L$ due to lower costs, increasing in $ P$ due to higher demand, 
            and convexly decreasing in $ G$ due to substitution by fossil fuel.
            Analogously, $ F$ is increasing in $ G, K, P$ and convexly decreasing in $ L$.
  \item $ Y$ is increasing and concave in all of $ E, K, P$.
\end{itemize}
We fulfil most of these by the following simple choices:
\begin{itemize}
  \item $ f_\text{diff.}( A, M) =  d ( M - m A)$
  \item $T=A$ (since $T$ is measured in carbon-equivalent degrees)
  \item $ f_\text{photos.}( A, L, T)
  			= ( l_0 -  l_T T)\sqrt{ A/ \Sigma}\, L$~\footnote{%
  			The exponent 1/2 for $ A$ in the fertilization term is larger but simpler than the choice of 0.3 in \cite{Anderies2013}.}
  \item $ f_\text{resp.}( L, T) = ( a_0 +  a_T T)  L$
  \item $ f_\text{deprec.}( K) =  k  K$
  \item $ W(L,P,Y) = (1 - i)  Y / P +  w_L  L$ with ecosystem services coefficient $w_L$
  \item $ f_\text{fert.}( W) = 2  p  W_P  W / ( W_P^2 +  W^2)$ 
  		with a maximum fertility of $p>0$ reached at the saturation well-being level $W_P>0$
  \item $ f_\text{mort.}( W) =  q/ W$ with mortality coefficient $q>0$
  \item $E_B = e_B B$ and $E_F = e_F F$ with combustion efficiencies $e_B,e_F>0$
\end{itemize}
The formulae for $B,F,Y$ are derived from the following economic submodel.

\subsubsection*{Two-sector economic submodel.}

We assume the global economy produces output using a global production function
\begin{eqnarray*}
     Y &=  f( P, K, L, G),
\end{eqnarray*}
using $P$ as a source of labour and $L,G$ as sources of energy.
In the full model, we assume larger population numbers lead to increasing globalization with overall positive effects on productivity,
hence we will aim at choosing an $f$ that has increasing returns to scale, i.e,
$ f(a P,a K,a L,a G) > a  f( P, K, L, G)$ for all $a>1$.
In the reduced model for pre-capitalistic societies, we will keep the more traditional assumption of constant returns to scale, i.e., $ f(a P,a K,a L,a G) = a  f( P, K, L, G)$ for all $a>1$.
This will influence our choice of elasticities (see below).
In order to be able to model substitution effects between the two different resource use flows $ B$ and $ F$, 
we need to distinguish the energy sector(s) from the rest of the economy (which we call the ``final'' sector).
A quite general modelling approach for doing this is to assume nested production functions
\begin{eqnarray*}
     Y &=  f( P, K, L, G) 
        =  f_Y( P_Y, K_Y, E_B, E_F),\\
     E_B &=  f_B( P_B, K_B, L),\\
     E_F &=  f_F( P_F, K_F, G)
\end{eqnarray*}
and determine the unknown labour and capital shares $ P_\cdot, K_\cdot$
by some form of social optimization or market mechanism.
Since this will in general lead to quite complicated expressions for $ Y, E_B, E_F$,
we make a number of strong simplifying and symmetry assumptions here
in order to get manageably simple formulae.

To reduce the number of independent factors in $f$, 
we treat the two energy forms as perfect substitutes, 
so that $Y = f_Y(P_Y,K_Y,E)$
with total energy input $E = E_B+E_F$.
Since energy is generally considered an input that cannot be substituted well by other factors,
the natural candidate to model the dependency of $Y$ on $E$
is not a CES production function but either a Cobb-Douglas or a Leontieff production function.
We choose the simpler, a Leontieff form, which amounts to prescribing a fixed
ratio of energy need per output that is independent of the other factors:
\begin{eqnarray*}
    Y &= y_E \min\{ E, g_Y(K_Y,P_Y) \},
\end{eqnarray*}
where $y_E > 0$ is an energy productivity factor (the inverse of the final sector's energy intensity).
We assume the standard Cobb-Douglas form for the relative substitutability of labour and capital:
\begin{eqnarray*}
    g_Y(K_Y,P_Y) &= b_Y K_Y^{\kappa_Y} P_Y^{\pi_Y}
\end{eqnarray*}
with productivity $b_Y>0$ and elasticities $0<\kappa_Y,\pi_Y<1$.
In each of the two forms of energy, we also assume the Cobb-Douglas form,
\begin{eqnarray*}
    E_B &= b_B K_B^{\kappa_B} P_B^{\pi_B} L^\lambda,\\
    E_F &= b_F K_F^{\kappa_F} P_F^{\pi_F} G^\gamma,
\end{eqnarray*}
with sectoral productivities $b_B,b_F>0$ and further elasticities $\kappa_\cdot, \pi_\cdot,\lambda,\gamma$.

Although the simplest assumption about the allocation
of labour and capital to the three production processes $f_Y,f_B,f_F$
would be to assume fixed shares, 
this would ignore the strong incentive to allocate the resources 
to the production of the more productive energy form, 
and to allocate the more resources to energy production the more productive the energy sector is compared to the final sector.
The next-best simple assumption is a social planner perspective that allocates
resources so as to maximize final output $Y$.
We prefer this to the alternative view of a competitive allocation via factor markets for two reasons:
(i) the latter view is more closely tied to the assumption of a specific economic system,
which is less plausible for the long time horizons we aim at,
and (ii) if markets are approximately perfect, they would lead to maximizing final output anyway.

To get this solution, we first assume the energy sector's inputs $K_E, P_E$ 
were known and solve the intra-energy-sector allocation problem
via the first-order conditions
\begin{eqnarray*}
    \partial E_B/\partial K_B &= \partial E_F/\partial K_F,\quad
    \partial E_B/\partial P_B = \partial E_F/\partial P_F
\end{eqnarray*}
under the constraints
\begin{eqnarray*}
    K_B + K_F + K_R &= K_E,\quad
    P_B + P_F + P_R = P_E.
\end{eqnarray*}
It turns out that this only leads to sufficiently simple expressions
if we assume that the labour elasticities $\pi_B,\pi_F$ of the two energy forms are equal,
and similarly for capital,
hence we put
$\kappa_{B,F}\equiv\kappa_E$ and $\pi_{B,F}\equiv\pi_E$ and get
\begin{eqnarray*}
    K_B &= X_B K_E / X_E,\quad
    K_F = X_F K_E / X_E,\\
    P_B &= X_B P_E / X_E,\quad
    P_F = X_F P_E / X_E,\\
    E_B &= X_B Z_E,\quad
    E_F = X_F Z_E,
\end{eqnarray*}
where
\begin{eqnarray*}
    X_B &= b_B^{\alpha_E} L^{\alpha_E\lambda},\quad
    X_F = b_F^{\alpha_E} G^{\alpha_E\gamma},
\end{eqnarray*}
\begin{eqnarray*}
    X_E &= X_B + X_F,\\
    Z_E &= K_E^{\kappa_E} P_E^{\pi_E} / X_E^{\kappa_E + \pi_E},\\    
    \alpha_E &= 1 / (1 - \kappa_E - \pi_E).
\end{eqnarray*}
Given $K_E, P_E$, we thus have
\begin{eqnarray*}
    E   &= X_E Z_E = K_E^{\kappa_E} P_E^{\pi_E} X_E^{1 / \alpha_E}.
\end{eqnarray*}
Since neither the energy nor the final sector are to have idle resources,
we must also have 
\begin{eqnarray*}
	E &= g_Y(K_Y,P_Y)= b_Y K_Y^{\kappa_Y} P_Y^{\pi_Y}.
\end{eqnarray*}
An optimal allocation between energy and final sector then requires
that no ``trade'' in capital or labour is profitable beween the two sectors,
which in view of the constraint $E = g_Y$ leads to the additional equation
\begin{eqnarray*}
    \frac{\partial g_Y / \partial K_Y}{\partial E / \partial K_E}
    &= \frac{\partial g_Y / \partial P_Y}{\partial E / \partial P_E},
\end{eqnarray*}
i.e.,
\begin{eqnarray*}
    \frac{\kappa_Y g_Y / K_Y}{\kappa_E E / K_E}
    &= \frac{\pi_Y g_Y / P_Y}{\pi_E E / P_E}
\end{eqnarray*}
which implies
\begin{eqnarray*}
    \frac{\kappa_Y K_E}{\kappa_E K_Y}
    &= \frac{\pi_Y P_E}{\pi_E P_Y}
    =: \beta.
\end{eqnarray*}
To find $\beta$, we solve
\begin{eqnarray*}
    0 &= E - g_Y\\
    &= \left(\frac{\beta\kappa_E K_Y}{\kappa_Y}\right)^{\kappa_E} \left(\frac{\beta\pi_E P_Y}{\pi_Y}\right)^{\pi_E} X_E^{1 / \alpha_E}
    - b_Y K_Y^{\kappa_Y} P_Y^{\pi_Y}
\end{eqnarray*}
and get
\begin{eqnarray*}
    \beta^{\kappa_E + \pi_E} &=
        b_Y \left(\frac{\kappa_Y}{\kappa_E}\right)^{\kappa_E} \left(\frac{\pi_Y}{\pi_E}\right)^{\pi_E} 
        K_Y^{\kappa_Y - \kappa_E} P_Y^{\pi_Y - \pi_E} X_E^{-1 / \alpha_E}.
\end{eqnarray*}
We note that this simplifies considerably if for each of the factors capital and labour,
either only one of the sectors requires it or both sectors have the same elasticity for it.
Since clearly a considerable amount of capital and labour are needed in both sectors,
we hence assume $\kappa_E = \kappa_Y =: \kappa$ and $\pi_E = \pi_Y =: \pi$.
We can now solve
\begin{eqnarray*}
    \frac{K_E}{K - K_E} &= \frac{P_E}{P - P_E} = \beta = (b_Y X_E^{-1 / \alpha_E})^{1 / (\kappa + \pi)},
\end{eqnarray*}
\begin{eqnarray*}
    K_E &= \frac{\beta}{1+\beta} K,\quad
    P_E  = \frac{\beta}{1+\beta} P,\\
    K_Y &= \frac{1}{1+\beta} K,\quad
    P_Y  = \frac{1}{1+\beta} P.
\end{eqnarray*}
Putting all of the above together, 
using $\eta = 1/(1+1/\beta)$ (the share of the energy sector) instead of $\beta$,
and introducing $\alpha = 1 / (1 - \kappa - \pi)$, $a_B = b_B^\alpha$ and $a_F = b_F^\alpha$,
we get
\begin{eqnarray*}
    X_B &= a_B L^{\alpha\lambda},\quad
    K_B  = \frac{X_B}X K_E,\quad
    P_B  = \frac{X_B}X P_E,\\   
    X_F &= a_F G^{\alpha\gamma},\quad
    K_F  = \frac{X_F}X K_E,\quad
    P_F  = \frac{X_F}X P_E,\\
    X    &= X_B + X_F,\quad
    \eta = \frac{1}{1 + (X^{1 / \alpha} / b_Y)^{1 / (\kappa + \pi)}},\\
    Z   &= K_E^\kappa P_E^\pi / X^{\kappa + \pi} = \eta^{\kappa + \pi} K^\kappa P^\pi / X^{\kappa + \pi},\\    
    E   &= X Z,\quad
    K_E  = \eta K,\quad
    P_E  = \eta P,\\ 
    Y   &= y_E E,\quad
    K_Y  = (1 - \eta) K,\quad
    P_Y  = (1 - \eta) P,\\
    Z' &= \left(1 + \frac{\left(a_B L^{\alpha\lambda} + a_F G^{\alpha\gamma}\right)^{\frac{1-\kappa-\pi}{\kappa+\pi}}}{b_Y^{1 / (\kappa + \pi)}}\right)^{-\kappa - \pi},\\
    E_B &= X_B Z 
    	=\textstyle \frac{a_B L^{\alpha\lambda} K^\kappa P^\pi}{\left(a_B L^{\alpha\lambda} + a_F G^{\alpha\gamma}\right)^{\kappa + \pi}}
    	Z', \\
    E_F &= X_F Z 
    	=\textstyle \frac{a_F G^{\alpha\gamma} K^\kappa P^\pi}{\left(a_B L^{\alpha\lambda} + a_F G^{\alpha\gamma}\right)^{\kappa + \pi}}
    	Z'.
\end{eqnarray*}

For the economy to have increasing returns to scale,
we choose elasticities that fulfil $\kappa + \pi + \min(\lambda,\gamma) > 1$.
A simple choice which is roughly in line with estimates of
labour and capital elasticities in the agricultural sector of many countries is
$\kappa=\pi=\lambda=\gamma=2/5$.
Then $\kappa+\pi=4/5$, $\alpha=5$, $\alpha\lambda=\alpha\gamma=2$,
and hence
\begin{eqnarray*}
    E_B &=\textstyle \frac{a_B L^2 (P K)^{2/5}}{\left(a_B L^2 + a_F G^2\right)^{4/5}}
    	\left(1 + \frac{\left(a_B L^2 + a_F G^2\right)^{1/4}}{b_Y^{5/4}}\right)^{-4/5}, \\
    E_F &=\textstyle \frac{a_F G^2 (P K)^{2/5}}{\left(a_B L^2 + a_F G^2\right)^{4/5}}
    	\left(1 + \frac{\left(a_B L^2 + a_F G^2\right)^{1/4}}{b_Y^{5/4}}\right)^{-4/5}.
\end{eqnarray*}
Finally, we assume that $b_Y^5 \gg a_B L^2 + a_F G^2$
so that the share of the energy sector $\eta$ (the large bracket) is $\approx 1$.
Note that as the ``energy'' sector in our model includes all of agriculture, 
a very large share of this sector is not too implausible.
We thus arrive at the simple approximation used in the model,
\begin{eqnarray*}
    B &= \frac{a_B}{e_B}\frac{L^2 (P K)^{2/5}}{(a_B L^2 + a_F G^2)^{4/5}}, \\
    F &= \frac{a_F}{e_F}\frac{G^2 (P K)^{2/5}}{(a_B L^2 + a_F G^2)^{4/5}}, \\
    Y &= y_E (e_B B + e_F F).
\end{eqnarray*}

For the pre-capitalistic variant of the model, we choose $\kappa=\lambda=3/10$ instead to get constant returns to scale.
Together with a fixed per capita capital of $K\propto P$, this gives equations \ref{eqn:B-PI} and \ref{eqn:Y-PI}.

\section{Parameter estimation}
\label{sec:par-est}

The available Earth surface area ($\Sigma$) has been identified with the Earth's current land surface area. The parametrization of the carbon cycle parameters ($C^\ast$, $C^\ast_\text{PI}$, $a_0$, $a_T$, $l_0$, $l_T$, $d$, $m$) occurred on the basis of the recent estimated of carbon stocks and flows by the International Panel on Climate Change \cite{Ciais2013}. The estimates of the demographic parameters ($p$, $W_P$, $q$) result from separately performed weighted least squares regressions of the modelled dependencies of fertility and mortality on well-being (equation (\ref{eqn:Pdot})), respectively. As input data we used estimates of various World Development Indicators for which country-wise, yearly data are available from the World Bank \cite{worldbank-wdi}.
The investment rate ($i$) has been estimated by averaging the global times series on ``gross capital formation'' by the World Bank \cite{worldbank-wdi}. A reasonable value for the capital depreciation rate ($k$) can be found in \cite{Nadiri1996}. Typical energy densities of biomass ($e_B$) and fossil fuels ($e_F$) are of comparable size \cite{McKendry2002}. The economic output per (primary) energy input has been estimated as the average of the inverse of the time series on ``energy intensity level of primary energy'' available from the World Bank \cite{worldbank-wdi}.

The subsequently introduced parameters $y_B$ and $b$ in the non-fossil scenario (section \ref{sec:non-fossil}) have been estimated using data on global population level, agricultural sector's value added to the gross world product and the contribution of harvesting to the ``Human Appropriation of Net Primary Production'' (HANPP) \cite{worldbank-wdi, Krausmann2013}.

\section{Bifurcation analysis}
\label{sec:bifurcations}

The rather low-dimensional complexity and the simple functional relationships (cf. equations (\ref{eqn:Ldot}) to (\ref{eqn:Kdot})) of the presented model facilitate the application of analysis techniques from dynamical systems theory, e.g. bifurcation analysis \cite{Kuznetsov1998}. Bifurcation analysis aims at a partition of a dynamical system's parameter space into regimes, such that within different regimes the system's state spaces are topologically non-equivalent, meaning different numbers or stabilities of the system's equilibria or limit cycles and hence a different asymptotic behaviour.

For this work we conducted a bifurcation analysis of the $(y_B,w_L)$-parameter-subspace of the two-dimensional $(L,P)$ submodel discussed in section \ref{sec:non-fossil}. The bifurcation diagram in figure \ref{fig:bifurcation-diagram}) shows a partition of the parameter space into five regimes for which the corresponding state spaces are topologically non-equivalent. The borders between the regimes correspond to codimension-1-bifurcations, while the blue points at their intersections indicate bifurcations of codimension 2.

Suppose the parameter values lie within the large reddish region in figure \ref{fig:bifurcation-diagram} for which the ``desert'' state is the only attractor of the system. When crossing the red curve above the blue square, the system undergoes a (local) \textit{fold} (or \textit{saddle-node}) bifurcation leading to the existence of an unstable (saddle) equilibrium and a stable (node) equilibrium in the dark green regime which hence facilitates a sustainable coexistence of humans with nature. Crossing the green curve gives rise to a (global) homoclinic bifurcation through which an unstable limit-cycle is created. However, this does not alter the set of attractors, hence the qualitative asymptotics remain unchanged. If the orange curve is transgressed from within the light green region, an \textit{Andronov-Hopf} bifurcation occurs. It is \textit{sub-critical} when the curve is crossed above the blue circle. In this case the unstable limit-cycle coalesces with the stable node, leaving an unstable node in the orange region. When the orange curve is crossed below the blue circle, the Andronov-Hopf bifurcation is \textit{super-critical}, meaning that a stable limit-cycle is born around the stable coexistence equilibrium which in turn becomes unstable. The yellow region hence features an attracting limit-cycle besides the stable desert state. The yellow bifurcation curve corresponds to a \textit{fold bifurcation of cycles} in which the two limit-cycles coalesce and vanish, leaving an unstable node in the orange region. Hence, in the orange regime the systems features a saddle point and an unstable node with $P>0$, which undergo a fold bifurcation when the red line is crossed from left to right below the blue square. In the orange and red regions the desert state is the only attractor, meaning that ultimately nature and society are doomed to collapse.

At the point marked by the blue square at which the fold, Andronov-Hopf and homoclinic bifurcation curves intersect, a so-called \textit{Bogdanov-Takens} bifurcation occurs. The point marked by the blue square at which the fold-of-cycles curve connects to the two branches of the Andronov-Hopf curve is referred to as a \textit{Bautin} (or \textit{generalized Hopf}) bifurcation.

Note that in figure \ref{fig:bifurcation-diagram} only the fold and Andronov-Hopf curves which correspond to \textit{local} bifurcations have been computed numerically, using the software \texttt{PyDSTool} \cite{Clewley2012}. As the tool is not able to detect \textit{global} bifurcations, the homoclinic and fold-of-cycles curves, whose existence is known from theory, are indicated only schematically.

\newpage
\section{Conditions for superexponential growth}
\label{sec:superexponential}

Due to several nonlinearities in our model, most quantities can show both sub- and superexponential growth or decay, in contrast to most basic purely economic growth models.

A quantity $x$ has a phase of superexponential growth whenever
$0 < d^2(\ln x)/dt^2 = (\ddot x x - \dot x^2) / x^2$.

For population $P$, we have 
$d(\ln P)/dt = \dot P/P = f(W) := \frac{2 W W_P}{W^2 + W_P^2}\,p - \frac{q}{W}$
and $f$ is negative if $0 < W < W_0$ (for some constant $W_0$), positive and increasing if $W_0 < W < W^\ast$, and positive and decreasing if $W^\ast<W$, where $0 < W_0 < W_P < W^\ast$.
Hence $P$ has superexponential growth iff 
either (i) $W_0 < W < W^\ast$ and $\dot W > 0$,
or (ii) $W^\ast<W$ and $\dot W<0$,
i.e., when well-being is moving towards the point where net reproduction is maximal.

For capital $K$, the condition is
\begin{eqnarray*}
	0 &< \ddot K K - \dot K^2\\
    &= K\,\frac{d}{dt}\left(i y_B (a_B L^2 + a_F G^2)^{1/5} (P K)^{2/5} - k_0 K\right) 
    	- \dot K^2\\
    &=\textstyle K\,\left(i y_E (a_B L^2 + a_F G^2)^{1/5} (P K)^{2/5}\right.\\
    			&\left.\quad\quad\times\left(
                	\frac{2 a_B L \dot L + 2 a_F G \dot G}{5(a_B L^2 + a_F G^2)}
                    + \frac{2 \dot P}{5 P}
                    + \frac{2 \dot K}{5 K}
                \right)\right.\\
    			&\left.\quad\quad- k_0 \dot K\right) 
    	- \dot K^2\\
    &= K\,\left((\dot K + k_0 K)\frac{2}{5} 
    			\left(
                	\frac{a_B L \dot L + a_F G \dot G}{a_B L^2 + a_F G^2}
                    + \frac{\dot P}{P}
                    + \frac{\dot K}{K}
                \right)\right.\\
    			&\left.\quad\quad- k_0 \dot K\right) 
    	- \dot K^2.
\end{eqnarray*}
If $\dot K > 0$, this condition is the more likely fulfilled
the smaller $\dot K$, $L$, $G$, and $P$,
and the larger $K$, $\dot L$, $\dot G$, $\dot P$, and $k_0$.
Hence a small $l_T$, $a_0$, $a_T$, $i$, $y_E$, $a_B$, $a_F$, $q$, and $q_P$,
a large $A$, $l_0$, $e_B$, $e_F$, and $p$,
and a $W\approx W_P$
tend to make a superexponential growth of $K$ more likely.

\end{document}